\documentclass[aps,pra,preprint,superscriptaddress]{revtex4-1}
\usepackage{amsmath}
\usepackage{amssymb}
\usepackage{graphicx}

\begin{document}

\title{Transfer of optical vortices in coherently prepared media}

\author{Hamid Reza Hamedi}
\email{hamid.hamedi@tfai.vu.lt}
\affiliation{Institute of Theoretical Physics and Astronomy, Vilnius University,
Saul\.etekio 3, Vilnius LT-10257, Lithuania}

\author{Julius Ruseckas}
\email{julius.ruseckas@tfai.vu.lt}
\affiliation{Institute of Theoretical Physics and Astronomy, Vilnius University,
Saul\.etekio 3, Vilnius LT-10257, Lithuania}

\author{Emmanuel Paspalakis}
\email{paspalak@upatras.gr }
\affiliation{Materials Science Department, School of Natural Sciences, University
of Patras, Patras 265 04, Greece}

\author{Gediminas Juzeli\={u}nas}
\email{gediminas.juzeliunas@tfai.vu.lt}
\affiliation{Institute of Theoretical Physics and Astronomy, Vilnius University,
Saul\.etekio 3, Vilnius LT-10257, Lithuania}

\begin{abstract}
We consider transfer of optical vortices between laser pulses carrying
orbital angular momentum (OAM) in a cloud of cold atoms characterized
by the $\Lambda$ configuration of the atom-light coupling. The atoms
are initially prepared in a coherent superposition of the lower levels,
creating a so-called phaseonium medium. If a single vortex beam initially
acts on one transition of the scheme, an extra laser beam is subsequently
generated with the same vorticity as that of the incident vortex beam.
The absorption of the incident probe beam takes place mostly at the
beginning of the atomic medium within the absorption length. The losses
disappear as the probe beam propagates deeper into the medium where
the atoms are transferred to their dark states. The method is extended
to a tripod atom-light coupling scheme and a more general $n+1$-level
scheme containing $n$ ground states and one excited state, allowing
for creating of multiple twisted light beams. We also analyze generation
of composite optical vortices in the scheme using a superposition
of two initial vortex beams and study lossless propagation of such
composite vortices.
\end{abstract}


\maketitle

\section{Introduction}

The coherent manipulation of pulse propagation through atomic ensembles
\cite{Harris-EIT-1997,Mazets-PhysRevA-2005,Hioe-PhysRevA-2008,CsesznegiPRA1997}
leads a plethora of important phenomena, such as electromagnetically
induced transparency (EIT) \cite{Harris-EIT-1997,Lukin-RevModPhys-2003,Fleischhauer-RevModPhys-2005,E.PaspalakisPRA2002}
and slow light propagation \cite{Harris-EIT-1997,Fleischhauer-RevModPhys-2005,Lukin-RevModPhys-2003,Juzeliunas-PhysRevLett-2004},
enhancement of optical nonlinearities \cite{Harris-PhysRevLett-1990,Deng-PhysRevA-1998,Wang-phys.rev.lett.2001,Kang-PhysRevLett-2003},
generation of matched pulses \cite{Harris-PhysRevLett-1994,Cerboneschi-PhysRevA-1995},
creation of a spinor slow light \cite{Unanyan2010,Ruseckas2011PRA,Bao2011PRA,Ruseckas-PhysRevA.87-2013,Lee2014},
formation of adiabatons \cite{Grobe-PhysRevLett-1994,Fleischhauer-PhysRevA-1996}
and optical solitons \cite{Wu-PhysRevLett-2004,Wu-OL-2004}. It has
been demonstrated that due to the EIT the light pulses not only could
be slowed down, but also stored by switching off the controlled beam
\cite{Phillips-PhysRevLett-2001,Chien-Nature-2001,Lukin-RevModPhys-2003,Lukin-Nature-2001,Juzeliunas2002}.
Therefore, the atomic system can be used as an optical memory for
transferring the quantum state of light to the matter and back to
the light \cite{Lukin-RevModPhys-2003,Hsiao-PhysRevLett-2018,Lijun-J.Opt-2017}.
By using extra energy levels and additional laser field one arrives
at more complex atom-light coupling schemes \cite{Unanyan98OC,Unanyan99PRA,Mazets-PhysRevA-2005,Knight-Opt,Ruseckas2011,Korsunsky-pra1999,Liu-PhysRevA-2004,Payne-PhysRevA-2002,Ruseckas-PhysRevA-2018,Bao-PhysRevA-2011,Hamedi-JPB-2017,PaspalakisPhysRevA2002CPM,PaspalakisPhysRevA2002multi}
which can provide more than one dark states and offer different directions
in studying of propagation effects in coherently driven atomic media.

Light can carry an orbital angular momentum (OAM) \cite{Allen1999,Miles-physToday-2004}.
Such light beams have helical (twisted) wavefronts that spiral along
the beam direction much like a corkscrew. The twisted light field
is characterized by a phase factor $e^{il\phi}$, where $\phi$ is
the azimuthal angle with respect to the beam axis and $l$ denotes
the vortex integer winding number (OAM number). When interacting with
atoms such optical vortex beams reveal a number of interesting effects,
including light-induced-torque \cite{Babiker-PhysRevLett1994,Lembessis-PhysRevA-2010},
atom vortex beams \cite{Lembessis-PhysRevA.89-2014}, entanglement
of OAM states of photon pairs \cite{Chen-PhysRevA-2008}, OAM-based
four-wave mixing \cite{Ding-OL-2012,WalkerPhysRevLett2012}, spatially
dependent optical transparency \cite{RadwellPhysRevLet2015,SharmaPhysRevA2017,Hamedi2018OE},
and the vortex slow light \cite{Dutton-PhysRevLett-2004,Ruseckas-PhysRevA-2007,Ruseckas2011,Hamedi-PhysRevA-2018}.
The twisted slow light \cite{Dutton-PhysRevLett-2004,Ruseckas-PhysRevA-2007,Ruseckas2011,Hamedi-PhysRevA-2018,Wang-PhysRevA-2008,Ruseckas-PhysRevA.-2011,Ruseckas-PhysRevA.87-2013}
gives additional possibilities in manipulation of the optical information
during the storage and retrieval of the slow light \cite{Pugatch-PhysRevLet-2004,Moretti-PhysRevA-2009}.

The previous studies on the EIT have concentrated on a situation where
the atoms are initially in their ground states, and the Rabi-frequency
of the probe field is much weaker than that of the control field.
It has been demonstrated that the OAM of the control vortex beam can
be transferred to the probe field in the tripod atom-light coupling
scheme during the storage and retrieval of the probe field \cite{Ruseckas-PhysRevA.-2011,Ruseckas2011}.
Without switching off and on of the control fields (hence without
storage and retrieval of the probe field), transfer of optical vortices
take place by applying a pair of weaker probe fields in the closed
loop double-$\Lambda$ \cite{Hamedi-PhysRevA-2018} or double-tripod
\cite{Ruseckas-PhysRevA.87-2013} schemes. 

In this paper, we demonstrate that the exchange of optical vortices
in non-closed loop structures is possible under the condition of weak
atom-light interaction in coherently prepared atomic media. To this
end, we analyze the interaction of multi-component laser pulses carrying
OAM propagating in multi-level atom-light coupling schemes with atoms
prepared in a coherent superposition of lower levels. Such a medium
has been named the phaseonium \cite{Scully1992,Scully-PhysRevLett-1985,Fleischhauer-PhysRevA-1992,EberlyPRL1996}.
We derive the basic equations describing the propagation of the coherent
laser pulses weakly interacting with atoms in multi-level configurations.
To elucidate the physical situation of exchange of OAMs, we begin
with a basic three-level configuration, the $\Lambda$ system, containing
only a pair of laser pulses. Subsequently we extend our model to more
complicated schemes involving additional laser pulses and additional
atomic levels. It is shown that the transfer of optical vortices is
also possible for the tripod system and a more general $n+1$-level
scheme. Furthermore we show that composite optical vortices can be
formed in the $\Lambda$ system when both probe fields are present
at entrance of the medium. 

The phaseonium medium proposed in this paper is based on a coherent
superposition of the ground states and can be realized experimentally
using the fractional or partial Stimulated Raman adiabatic passage
(STIRAP) \cite{Vitanov-RevModPhys-2017}. The generation of a quantum
superposition of ground states in a robust and controlled way is known
to be possible in a four-state tripod system by using a sequence of
three laser pulses \cite{Unanyan98OC,Unanyan99PRA}. Such a technique
is based on the existence of two degenerate dark states and their
interaction. The mixing of the dark states can be controlled by changing
the relative delay of the pulses, and thus an arbitrary superposition
state can be created. This method for creation of coherent superpositions
can be generalized to $N$ level schemes.

The method described here for transfer of optical vortices may find
application for creation of structured light by another light \cite{Dutton-PhysRevLett-2004}.
Using our method one could create a vortex at a wevelength for which
it is not possible to do it directly with standard optics (e.g. far
infrared or UV) \cite{Chaitanya2016}. In addition, the transfer of
vortices is a possible tool for manipulation of information encoded
into OAM of light. 

\section{The three-level $\Lambda$ system}

\begin{figure}
\includegraphics[width=0.5\columnwidth]{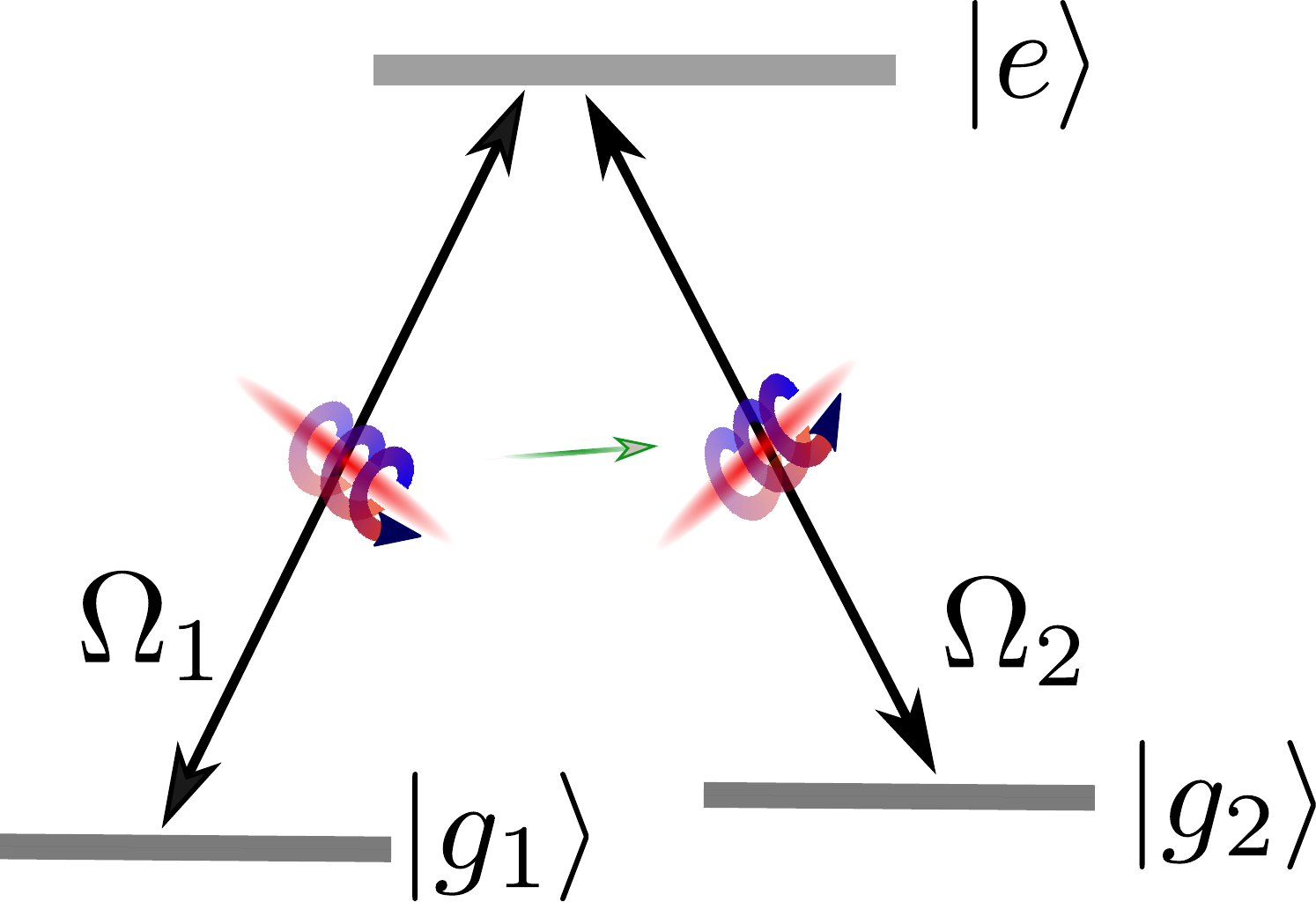}
\caption{Schematic diagram of the three-level $\Lambda$ quantum system containing
an upper state $|e\rangle$ and lower levels $|g_{1}\rangle$ and
$|g_{2}\rangle$ interacting with two Rabi-frequencies $\Omega_{1}$
and $\Omega_{2}$. }
\label{fig:1}
\end{figure}
Let us first consider the $\Lambda$ scheme for the transfer of optical
vortices, illustrated in Fig.~\ref{fig:1}. Specifically, we study
the propagation of two laser pulses with the Rabi-frequencies $\Omega_{1}$
and $\Omega_{2}$ (pulse pair) in a medium consisting of atoms in
the three-level $\Lambda$ configuration of the atom-light coupling.
The two atomic lower states $|g_{1}\rangle$ and $|g_{2}\rangle$
are coupled to an excited state $|e\rangle$ via the two light fields.
The Hamiltonian for such a system reads in the appropriate rotating
frame and in the interaction picture:
\begin{equation}
H_{\mathrm{\Lambda}}=\Omega_{1}|g_{1}\rangle\langle e|+\Omega_{2}|g_{2}\rangle\langle e|+\mathrm{H.c.}\,.\label{eq:1}
\end{equation}

The dynamics of the pulse pair $\Omega_{1}$ and $\Omega_{2}$ and
two atomic coherences $\rho_{g_{1}e}$ and $\rho_{g_{2}e}$ are described
by the Maxwell\textendash Bloch equations (MBE) for an open system
\begin{align}
\dot{\rho}_{g_{1}e}=&i(\delta_{1}+i\gamma_{eg_{1}})\rho_{g_{1}e}-i\Omega_{1}(\rho_{ee}-\rho_{g_{1}g_{1}})+i\Omega_{2}\rho_{g_{1}g_{2}}\,,\label{eq:2}\\
\dot{\rho}_{g_{2}e}=&i\left(\delta_{2}+i\gamma_{eg_{2}}\right)\rho_{g_{2}e}-i\Omega_{2}(\rho_{ee}-\rho_{g_{2}g_{2}})+i\Omega_{1}\rho_{g_{2}g_{1}}\,,\label{eq:3}
\end{align}
and
\begin{align}
\frac{\partial\Omega_{1}}{\partial z} & +c^{-1}\frac{\partial\Omega_{1}}{\partial t}=i\frac{\alpha_{1}\gamma_{eg_{1}}}{2L}\rho_{g_{1}e}\,,\label{eq:4}\\
\frac{\partial\Omega_{2}}{\partial z} & +c^{-1}\frac{\partial\Omega_{2}}{\partial t}=i\frac{\alpha_{2}\gamma_{eg_{2}}}{2L}\rho_{g_{2}e}\,,\label{eq:5}
\end{align}
where $\alpha_{_{1}}$ and $\alpha_{2}$ are the optical depths of
both laser pulses $\Omega_{1}$ and $\Omega_{2}$, $L$ denotes the
optical length of the medium, and $\gamma_{eg_{1}}$ and $\gamma_{eg_{2}}$
are the rates of decay from the excited state $|e\rangle$ to lower
states $|g_{1}\rangle$ and $|g_{2}\rangle$, respectively. We have
defined the detunings as $\delta_{1}=\omega_{eg_{1}}-\omega_{1}$
and $\delta_{2}=\omega_{eg_{2}}-\omega_{2}$, where $\omega_{eg_{1}}$
and $\omega_{eg_{2}}$ are the frequencies of the transitions $|g_{1}\rangle\leftrightarrow|e\rangle$
and $|g_{2}\rangle\leftrightarrow|e\rangle$, respectively, while
$\omega_{1}$ and $\omega_{2}$ represent the central frequencies
of the probe beams. We have disregarded the diffraction terms containing
the transverse derivatives $\left(2k_{1}\right)^{-1}\nabla_{\perp}^{2}\Omega_{1}$
and $\left(2k_{2}\right)^{-1}\nabla_{\perp}^{2}\Omega_{2}$ in the
Maxwell equations (\ref{eq:4}) and (\ref{eq:5}), where $k_{1}=\omega_{1}/c$
and $k_{2}=\omega_{2}/c$ are the central wave vectors of the first
and second beams. One can evaluate these terms as $\nabla_{\perp}^{2}\Omega_{1(2)}\sim w^{-2}\Omega_{1(2)}$,
where $w$ represents a characteristic transverse dimension of the
laser beams. This can be a width of the vortex core if the beam carries
an optical vortex or a characteristic width of the beam if it has
no vortex. Consequently the change of the phase of the probe beams
due to the diffraction term after passing the medium is estimated
to be $L/2kw^{2}$, where $L$ is the length of the atomic cloud,
where $k\approx k_{1(2)}$. The phase change $L/2kw^{2}$ can be neglected
when the sample length $L$ is not too large, $L\lambda/w^{2}\ll\pi$,
where $\lambda=2\pi/k$ is an optical wavelength. For example, by
taking the length of the atomic cloud to be $L=100\,\mathrm{\mu m}$,
the characteristic transverse dimension of the beams $w=20\,\mathrm{\mu m}$
and the wavelength $\lambda=1\,\mathrm{\mu m}$, we obtain $L\lambda/w^{2}=0.25$.
Under these conditions the diffraction terms do not play a significant
role and we can drop it out in Eqs.~(\ref{eq:4}) and (\ref{eq:5}).

Let us assume that the atoms are initially in a superposition of both
lower levels (the phaseonium medium)
\begin{equation}
|\psi(0)\rangle=c_{1}|g_{1}\rangle+c_{2}|g_{2}\rangle.\label{eq:atom-state1}
\end{equation}
We consider a weak atom-light interaction where $|\Omega_{1}|,|\Omega_{2}|\ll\gamma_{eg_{1}},\gamma_{eg_{2}}$.
Then, to the first order one has $\rho_{ee}\approx0$, $\rho_{g_{1}g_{1}}\approx|c_{1}|^{2}$,
$\rho_{g_{2}g_{2}}\approx|c_{2}|^{2}$ and $\rho_{g_{1}g_{2}}\approx c_{1}c_{2}^{*}$,
giving the following the steady state solutions for the coherences
$\rho_{g_{1}e}$ and $\rho_{g_{2}e}$:
\begin{align}
\rho_{g_{1}e}=&-\frac{|c_{1}|^{2}\Omega_{1}+c_{1}c_{2}^{*}\Omega_{2}}{\delta_{1}+i\gamma_{eg_{1}}},\label{eq:6}\\
\rho_{g_{2}e}=&-\frac{c_{1}^{*}c_{2}\Omega_{1}+|c_{2}|^{2}\Omega_{2}}{\delta_{2}+i\gamma_{eg_{2}}}.\label{eq:7}
\end{align}
The first-order approximation is valid when $|\rho_{g_{1}e}|,|\rho_{g_{2}e}|\ll1$.
Otherwise we cannot assume that $\rho_{g_{1}g_{1}}$ and $\rho_{g_{1}g_{2}}$
are not changing during the propagation of light. 

Substituting Eqs.~(\ref{eq:6}) and (\ref{eq:7}) into the Maxwell
equations (\ref{eq:4}) and (\ref{eq:5}) one arrives at the following
coupled equations for the propagation of the pulse pair \cite{PaspalakisPhysRevA2002multi}
\begin{align}
\frac{\partial\Omega_{1}}{\partial z}=&-i\beta_{1}(|c_{1}|^{2}\Omega_{1}+c_{1}c_{2}^{*}\Omega_{2}),\label{eq:8}\\
\frac{\partial\Omega_{2}}{\partial z}=&-i\beta_{2}(c_{1}^{*}c_{2}\Omega_{1}+|c_{2}|^{2}\Omega_{2}),\label{eq:9}
\end{align}
where
\begin{equation}
\beta_{a}=\frac{\alpha_{a}\gamma_{eg_{a}}}{2L(\delta_{a}+i\gamma_{eg_{a}})},\label{eq:10}
\end{equation}
with $a=1,2$.

The second laser field is assumed to be zero $\Omega_{2}(0)=0$ at
the entrance $z=0$, while $\Omega_{1}(0)=\Omega$. Under these conditions
the solutions to Eqs. (\ref{eq:8}) and (\ref{eq:9}) read
\begin{align}
\Omega_{1}(z)=&\frac{\Omega}{X_{2}}(\beta_{1}|c_{1}|^{2}e^{-iX_{2}z}+\beta_{2}|c_{2}|^{2}),\label{eq:11}\\
\Omega_{2}(z)=&\frac{\Omega}{X_{2}}c_{1}^{*}c_{2}\beta_{2}(e^{-iX_{2}z}-1),\label{eq:12}
\end{align}
where 
\begin{equation}
X_{2}=\beta_{1}|c_{1}|^{2}+\beta_{2}|c_{2}|^{2}.\label{eq:13-1}
\end{equation}
Up to now no assumption has been made made concerning the spatial
profile of the laser fields. We take now that the incident beam $\Omega_{1}$
has an optical vortex 
\begin{equation}
\Omega_{1}(0)=\Omega=|\Omega|e^{il\phi},\label{eq:15}
\end{equation}
where $l$ is the orbital angular momenta along the propagation axis
$z$, and $\phi$ is the azimuthal angle. For a doughnut Laguerre-Gaussian
(LG) beam the transverse profile reads
\begin{equation}
|\Omega|=\varepsilon(\frac{r}{w})^{|l|}e^{-r^{2}/w^{2}},\label{eq:16}
\end{equation}
where $r$ describes a cylindrical radius, $w$ is a beam waist, and
$\varepsilon$ represents the strength of the vortex beam. According
to Eqs.~(\ref{eq:11})-(\ref{eq:15}), the generated pulse beam $\Omega_{2}(z)\sim e^{il\phi}$
acquires the same phase as the first vortex beam. Therefore the laser
beam $\Omega_{1}$ transfers its vortex to the generated beam $\Omega_{2}$. 

\begin{figure}
\includegraphics[width=0.5\columnwidth]{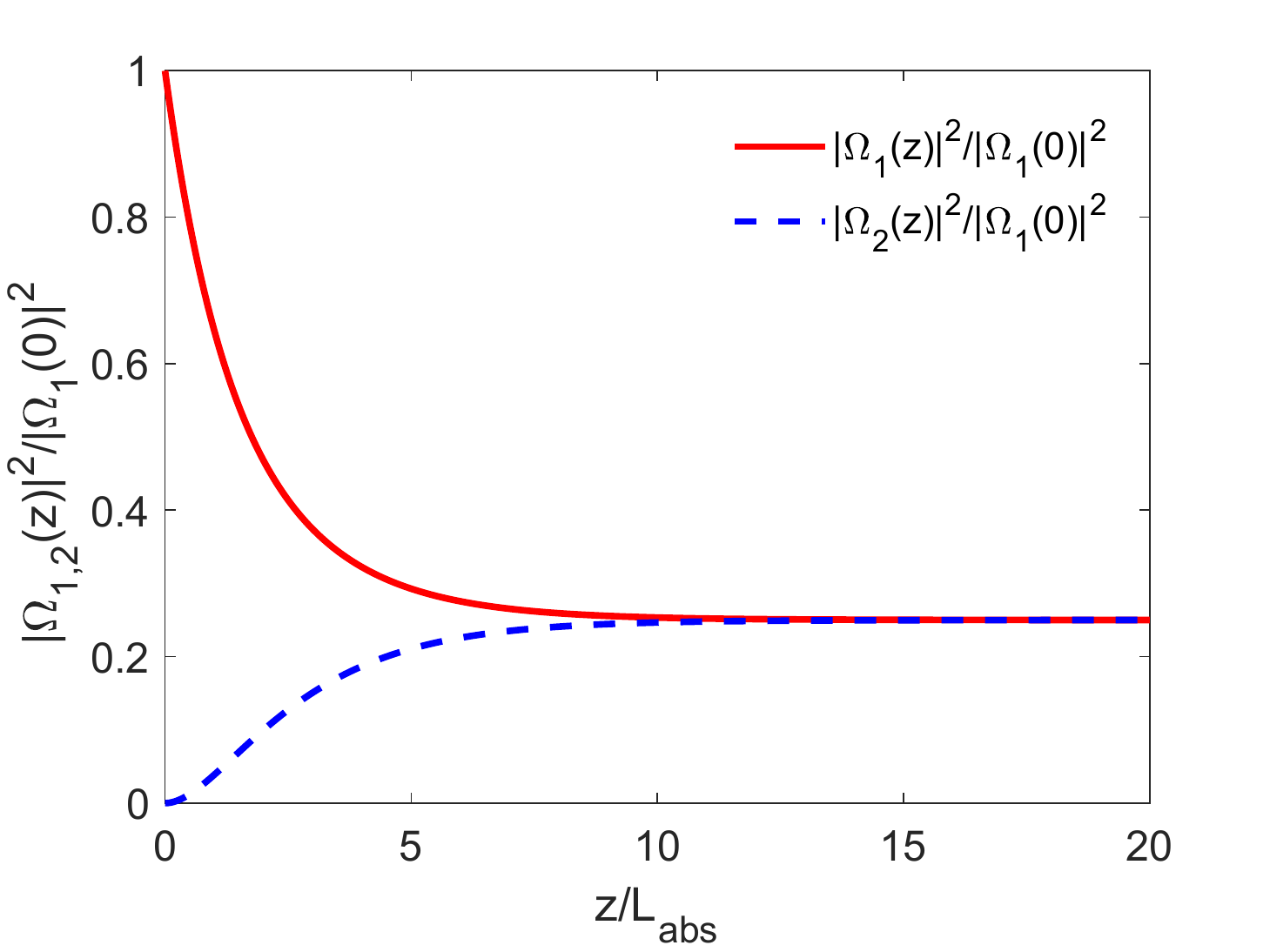}
\caption{Dependence of the dimensionless intensities of the light fields $|\Omega_{1}(z)|^{2}/|\Omega_{1}(0)|^{2}$
and $|\Omega_{2}(z)|^{2}/|\Omega_{1}(0)|^{2}$ given in Eqs.~(\ref{eq:11})
and (\ref{eq:12}) on the dimensionless distance $z/L_{abs}$ for
$c_{1}=c_{2}=\frac{1}{\sqrt{2}}$, $\delta_{1}=\delta_{2}=0$ and
$\alpha=20$.}
\label{fig:2}
\end{figure}

Equations~(\ref{eq:11}) and (\ref{eq:12}) show that both light
beams experience losses during their propagation. Yet the losses appear
only at the entrance of the medium before the EIT is established for
both fields. To simplify the discussion, let us take $\alpha_{1}=\alpha_{2}=\alpha$
and $\gamma_{eg_{1}}=\gamma_{eg_{2}}=\gamma$, and consider a situation
where both laser fields $\Omega_{1}$ and $\Omega_{2}$ are in an
exact resonance with the corresponding atomic transitions ($\delta_{1}=\delta_{2}=0$).
Then Eqs.~(\ref{eq:10}) and (\ref{eq:13-1}) lead to $\beta_{1}=\beta_{2}=X_{2}=\frac{1}{2iL_{abs}}$,
where $L_{abs}=L/\alpha$ is the absorption length. If optical density
of the resonant medium is sufficiently large $\alpha\gg1$, the absorption
length constitutes a fraction of the whole medium $L_{abs}\ll L$.
For the distances $z$ exceeding the absorption length $z\gg L_{abs}$
both exponential terms vanish in Eqs.~(\ref{eq:11}) and (\ref{eq:12})
and the EIT is established leading to lossless propagation of both
fields. Using Eqs.~(\ref{eq:15}) and (\ref{eq:16}) we get
\begin{align}
\Omega_{1}(z\gg L_{abs})=&\varepsilon(\frac{r}{w})^{|l|}e^{-r^{2}/w^{2}}(1-|c_{1}|^{2})e^{il\phi},\label{eq:16-1}\\
\Omega_{2}(z\gg L_{abs})=&-\varepsilon(\frac{r}{w})^{|l|}e^{-r^{2}/w^{2}}c_{1}^{*}c_{2}e^{il\phi}.\label{eq:16-2}
\end{align}
In this way, the beams experience no absorption loss for large propagation
distances $z\gg L_{abs}$. This is illustrated in Fig.~\ref{fig:2}
showing the dependence of the intensities $|\Omega_{1}(z)|^{2}/|\Omega_{1}(0)|^{2}$
and $|\Omega_{2}(z)|^{2}/|\Omega_{1}(0)|^{2}$ given by Eqs.~(\ref{eq:11})
and (\ref{eq:12}) on the dimensionless distance $z/L_{abs}$ for
the resonance case $\delta_{1}=\delta_{2}=0$ and $\alpha=20$. Although
initially at the beginning of the atomic medium losses occur, going
deeper through the medium the losses disappear as the atoms go to
their dark state \cite{Fleischhauer-RevModPhys-2005}
\begin{equation}
D(z\gg L_{abs})=\frac{\Omega_{2}(z\gg L_{abs})|g_{1}\rangle-\Omega_{1}(z\gg L_{abs})|g_{2}\rangle}{\sqrt{\Omega_{1}^{2}(z\gg L_{abs})+\Omega_{2}^{2}(z\gg L_{abs})}}.\label{eq:D}
\end{equation}

Let us investigate how sensitive is the proposed method for transferring
of optical vortices to errors in the amplitudes and the phases of
the superpositions. The sensitivity of system to the errors is given
by the derivative of the fields in the output given by Eqs.~(\ref{eq:11})
and (\ref{eq:12}) with respect to the coefficients $c_{1}$ and $c_{2}$.
Assuming $\beta_{1}=\beta_{2}=\beta=\frac{1}{2iL_{abs}}$ and using
the fact $|c_{2}|=\sqrt{1-|c_{1}|^{2}}$, Eqs.~(\ref{eq:11}) and
(\ref{eq:12}) can be rewritten as
\begin{align}
\Omega_{1}(z)=&\Omega+\Omega(e^{-i\frac{z}{2iL_{abs}}}-1)|c_{1}|^{2},\label{eq:11-1}\\
\Omega_{2}(z)=&\Omega(e^{-i\frac{z}{2iL_{abs}}}-1)|c_{1}|\sqrt{1-|c_{1}|^{2}}e^{i\phi_{c}},\label{eq:12-1}
\end{align}
where $\phi_{c}=\phi_{c_{2}}-\phi_{c_{1}}$ is the relative phase
of coefficients $c_{1}$ and $c_{2}$. The relative phase of the coefficients
$c_{1}$ and $c_{2}$ appears only in Eq.~(\ref{eq:12-1}). Calculating
the derivative of the fields given by Eqs.~(\ref{eq:11-1}) and (\ref{eq:12-1})
with respect to the amplitude $|c_{1}|$ as well as the relative phase
$\phi_{c}$, gives
\begin{align}
\frac{\partial\Omega_{1}(z)}{\partial|c_{1}|}=&2\Omega|c_{1}|(e^{-i\frac{z}{2iL_{abs}}}-1),\label{eq:derivative1}\\
\frac{\partial\Omega_{2}(z)}{\partial|c_{1}|}=&\Omega e^{i\phi_{c}}(e^{-i\frac{z}{2iL_{abs}}}-1)\frac{1-2|c_{1}|^{2}}{\sqrt{1-|c_{1}|^{2}}},\label{eq:derivative2}\\
\frac{\partial\Omega_{2}(z)}{\partial\phi_{c}}=&i\Omega_{2}(z).\label{eq:derivative3}
\end{align}
Equations~(\ref{eq:derivative1})-(\ref{eq:derivative3}) show that
the proposed method is not very sensitive to the errors in the coefficients.
This can be also seen from Eqs.~(\ref{eq:16-1}) and (\ref{eq:16-2})
for $z\gg L_{abs}$. It is clear that the ratio $|\Omega_{1}|/|\Omega_{2}|$
is proportional to $|c_{2}|/|c_{1}|$. Errors in the amplitudes will
change this ratio, and consequently, the intensity of the transferred
vortex. The intensity does not depend on the phases of the superpositions.
Only the relative phase of the coefficients $c_{1}$ and $c_{2}$
enters into Eq.~(\ref{eq:12}) or Eq.~(\ref{eq:16-2}) and changes
the global phase of the second field $\Omega_{2}(z)$. 

\section{The tripod system}

\begin{figure}
\includegraphics[width=0.5\columnwidth]{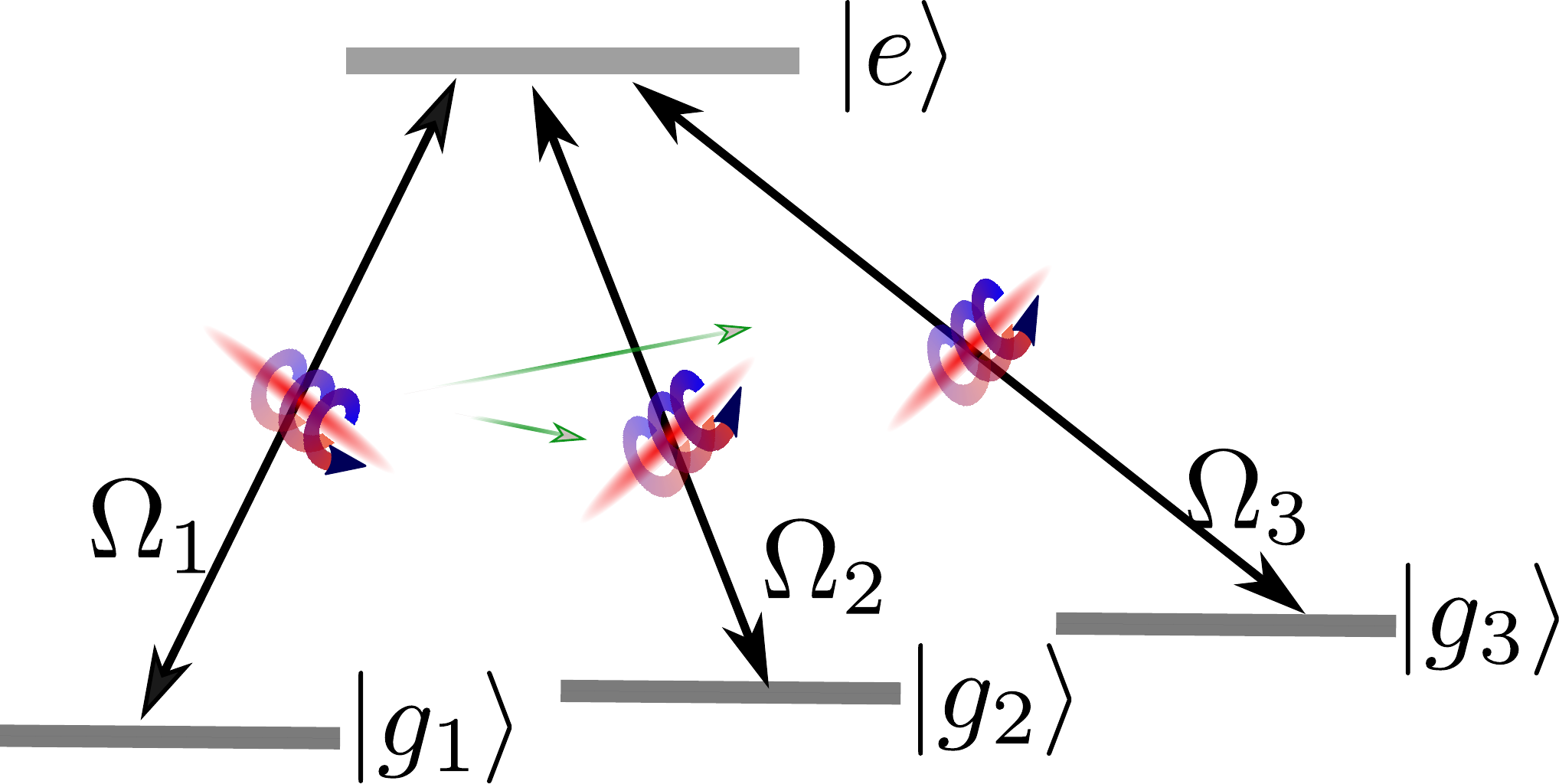}
\caption{Schematic diagram of the four-level tripod quantum system containing
an upper state $|e\rangle$ and lower levels $|g_{1}\rangle$, $|g_{2}\rangle$
and $|g_{3}\rangle$ interacting with three Rabi-frequencies $\Omega_{1}$,
$\Omega_{2}$ and $\Omega_{3}$.}
\label{fig:3}
\end{figure}
Consider next the propagation of three light pulses through a medium
consisting of atoms with a tripod level structure shown in Fig.~\ref{fig:3}.
Previously the tripod scheme was employed when studying the transfer
of optical vortices \cite{Ruseckas-PhysRevA.-2011,Ruseckas2011}.
It was shown that the transfer of optical vortices from a control
field of larger intensity to a probe field of weaker intensity in
tripod scheme was only possible through switching off and on of the
control laser beams \cite{Ruseckas2011,Ruseckas-PhysRevA.-2011}.
In the following, we demonstrate that the exchange of optical vortices
in tripod scheme is possible even without switching off and on of
the control beams, if one uses a coherently prepared atomic system. 

In the tripod scheme an excited state $|e\rangle$ is coupled to three
lower levels $|g_{1}\rangle$, $|g_{2}\rangle$ and $|g_{3}\rangle$
through three laser pulses $\Omega_{1}$, $\Omega_{2}$ and $\Omega_{3}$,
respectively. The Hamiltonian for the tripod scheme reads in the interaction
representation

\begin{equation}
H_{\mathrm{T}}=\Omega_{1}|g_{1}\rangle\langle e|+\Omega_{2}|g_{2}\rangle\langle e|+\Omega_{3}|g_{3}\rangle\langle e|+\mathrm{H.c.}\,.\label{eq:17}
\end{equation}
The MBEs describing the evolution of system can be written as
\begin{align}
\dot{\rho}_{g_{1}e}=&i(\delta_{1}+i\gamma_{eg_{1}})\rho_{g_{1}e}-i\Omega_{1}(\rho_{ee}-\rho_{g_{1}g_{1}})+i\Omega_{2}\rho_{g_{1}g_{2}}+i\Omega_{3}\rho_{g_{1}g_{3}}\,,\label{eq:18}\\
\dot{\rho}_{g_{2}e}=&i\left(\delta_{1}+i\gamma_{eg_{2}}\right)\rho_{g_{2}e}-i\Omega_{2}(\rho_{ee}-\rho_{g_{2}g_{2}})+i\Omega_{1}\rho_{g_{2}g_{1}}+i\Omega_{3}\rho_{g_{2}g_{3}}\,,\label{eq:19}\\
\dot{\rho}_{g_{3}e}=&i\left(\delta_{1}+i\gamma_{eg_{3}}\right)\rho_{g_{3}e}-i\Omega_{3}(\rho_{ee}-\rho_{g_{3}g_{3}})+i\Omega_{1}\rho_{g_{3}g_{1}}+i\Omega_{2}\rho_{g_{3}g_{2}}\,,\label{eq:20}
\end{align}
and
\begin{align}
\frac{\partial\Omega_{1}}{\partial z}+c^{-1}\frac{\partial\Omega_{1}}{\partial t}&=i\frac{\alpha_{1}\gamma_{eg_{1}}}{2L}\rho_{g_{1}e}\,,\label{eq:21}\\
\frac{\partial\Omega_{2}}{\partial z}+c^{-1}\frac{\partial\Omega_{2}}{\partial t}&=i\frac{\alpha_{2}\gamma_{eg_{2}}}{2L}\rho_{g_{2}e}\,,\label{eq:22}\\
\frac{\partial\Omega_{3}}{\partial z}+c^{-1}\frac{\partial\Omega_{3}}{\partial t}&=i\frac{\alpha_{3}\gamma_{eg_{3}}}{2L}\rho_{g_{3}e}\,,\label{eq:23}
\end{align}
where the diffraction terms have been neglected in the Maxwell equations
(\ref{eq:21}) and (\ref{eq:23}), like for the $\Lambda$ scheme. 

The atoms are initially prepared in a superposition of three lower
states
\begin{equation}
|\psi(0)\rangle=\text{c}_{1}|g_{1}\rangle+c_{2}|g_{2}\rangle+c_{3}|g_{3}\rangle.\label{eq:atom-state2}
\end{equation}
For a sufficiently weak atom-light interaction, $|\Omega_{j}|\ll\gamma_{eg_{j}}$,
we can approximate $\rho_{ee}\approx0$, $\rho_{g_{1}g_{1}}\approx|c_{1}|^{2}$,
$\rho_{g_{2}g_{2}}\approx|c_{2}|^{2}$, $\rho_{g_{3}g_{3}}\approx|c_{3}|^{2}$,
$\rho_{g_{1}g_{2}}\approx c_{1}c_{2}^{*}$ $\rho_{g_{1}g_{3}}\approx c_{1}c_{3}^{*}$
and $\rho_{g_{2}g_{3}}\approx c_{2}c_{3}^{*}$, giving the following
steady state equations for the atomic coherences
\begin{align}
\rho_{g_{1}e}=&-\frac{|c_{1}|^{2}\Omega_{1}+c_{1}c_{2}^{*}\Omega_{2}+c_{1}c_{3}^{*}\Omega_{3}}{\delta_{1}+i\gamma_{eg_{1}}},\label{eq:24}\\
\rho_{g_{2}e}=&-\frac{c_{1}^{*}c_{2}\Omega_{1}+|c_{2}|^{2}\Omega_{2}+c_{2}c_{3}^{*}\Omega_{3}}{\delta_{2}+i\gamma_{eg_{2}}},\label{eq:25}\\
\rho_{g_{3}e}=&-\frac{c_{1}^{*}c_{3}\Omega_{1}+c_{2}^{*}c_{3}\Omega_{2}+|c_{3}|^{2}\Omega_{3}}{\delta_{3}+i\gamma_{eg_{3}}}.\label{eq:26}\\
\end{align}
Substituting Eqs.~(\ref{eq:24})-(\ref{eq:26}) into the Maxwell
equations~(\ref{eq:21})-(\ref{eq:23}) and assuming that at the
entrance ($z=0$) there is a single light beam $\Omega_{1}(0)=\Omega$,
$\Omega_{2}(0)=0$, and $\Omega_{3}(0)=0$, we obtain \cite{PaspalakisPhysRevA2002CPM,PaspalakisPhysRevA2002multi}
\begin{align}
\Omega_{1}(z)=&\frac{\Omega}{X_{3}}(\beta_{1}|c_{1}|^{2}e^{-iX_{3}z}+\beta_{2}|c_{2}|^{2}+\beta_{3}|c_{3}|^{2}),\label{eq:27}\\
\Omega_{2}(z)=&\frac{\Omega}{X_{3}}c_{1}^{*}c_{2}\beta_{2}(e^{-iX_{3}z}-1),\label{eq:28}\\
\Omega_{3}(z)=&\frac{\Omega}{X_{3}}c_{1}^{*}c_{3}\beta_{3}(e^{-iX_{3}z}-1),\label{eq:29}
\end{align}
where
\begin{equation}
X_{3}=\beta_{1}|c_{1}|^{2}+\beta_{2}|c_{2}|^{2}+\beta_{3}|c_{3}|^{2},\label{eq:x3}
\end{equation}
and $\beta_{a}$ is defined by Eq.~(\ref{eq:10}), with $a=1,2,3$.
Considering again the first laser pulse $\Omega_{1}$ initially carries
an optical vortex (defined by Eqs.~(\ref{eq:15}) and (\ref{eq:16})),
two vortex beams $\Omega_{2}\sim e^{il\phi}$ and $\Omega_{3}\sim e^{il\phi}$
are generated with the same vorticity as the first laser pulse $\Omega_{1}\sim e^{il\phi}$. 

\begin{figure}
\includegraphics[width=0.5\columnwidth]{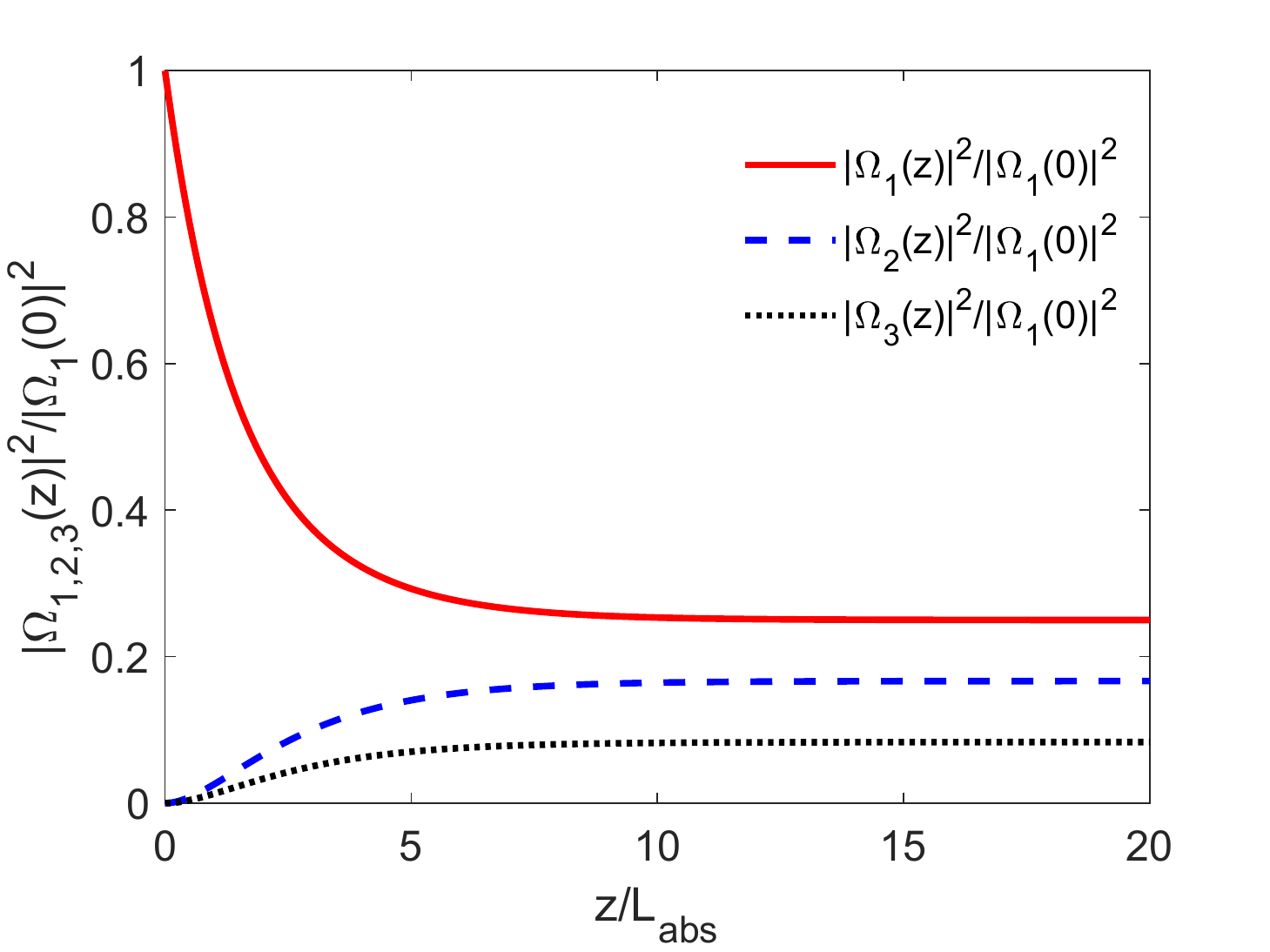}
\caption{Dependence of the dimensionless intensities $|\Omega_{1}(z)|^{2}/|\Omega_{1}(0)|^{2}$,
$|\Omega_{2}(z)|^{2}/|\Omega_{1}(0)|^{2}$ and $|\Omega_{3}(z)|^{2}/|\Omega_{1}(0)|^{2}$
given by Eqs.~(\ref{eq:27})-(\ref{eq:29}) on the dimensionless
distance $z/L_{abs}$ for $c_{1}=\frac{1}{\sqrt{2}}$, $c_{2}=\frac{1}{\sqrt{3}}$,
$c_{2}=\frac{1}{\sqrt{6}}$, $\delta_{1}=\delta_{2}=\delta_{3}=0$
and $\alpha=20$.}
\label{fig:4}
\end{figure}

Using Eqs.~(\ref{eq:15})-(\ref{eq:16}) and (\ref{eq:27})-(\ref{eq:29}),
one gets for sufficiently large $z$ ($z\gg L_{abs}$) under the resonance
condition $\delta_{1}=\delta_{2}=\delta_{3}=0$ and assuming that
$\alpha_{1}=\alpha_{2}=\alpha_{3}=\alpha$, $\gamma_{eg_{1}}=\gamma_{eg_{2}}=\gamma_{eg_{3}}=\gamma$:
\begin{align}
\Omega_{1}(z\gg L_{abs})=&\varepsilon(\frac{r}{w})^{|l|}e^{-r^{2}/w^{2}}(1-|c_{1}|^{2})e^{il\phi},\label{eq:29-1}\\
\Omega_{2}(z\gg L_{abs})=&-\varepsilon(\frac{r}{w})^{|l|}e^{-r^{2}/w^{2}}c_{1}^{*}c_{2}e^{il\phi},\label{eq:29-2}\\
\Omega_{3}(z\gg L_{abs})=&-\varepsilon(\frac{r}{w})^{|l|}e^{-r^{2}/w^{2}}c_{1}^{*}c_{3}e^{il\phi}.\label{eq:29-3}
\end{align}
Thus the lossless propagation of generated vortex beams takes place
at distances exceeding the absorption length $L_{abs}$ as illustrated
in Fig.~\ref{fig:4}. In that case the atomic systems goes to a superposition
of two dark states \cite{Unanyan98OC,Unanyan99PRA,PRA-tripod,Ruseckas-2005-tripod}
\begin{align}
D_{1}(z\gg L_{abs})=&\frac{\Omega_{3}(z\gg L_{abs})|g_{1}\rangle
-\Omega_{1}(z\gg L_{abs})|g_{3}\rangle}{\sqrt{\Omega_{1}^{2}(z\gg L_{abs})
+\Omega_{2}^{2}(z\gg L_{abs})}},\label{eq:D-1}\\
D_{2}(z\gg L_{abs})=&\left[
\Omega_{1}(z\gg L_{abs})\Omega_{2}(z\gg L_{abs})|g_{1}\rangle
+\Omega_{2}(z\gg L_{abs})\Omega_{3}(z\gg L_{abs})|g_{3}\rangle\right.\nonumber\\
&\left.-\left(\Omega_{1}^{2}(z\gg L_{abs})+\Omega_{3}^{2}(z\gg L_{abs})\right)|g_{2}\rangle\right]\nonumber\\
&\times\left[\left(\Omega_{1}^{2}(z\gg L_{abs})+\Omega_{3}^{2}(z\gg L_{abs})\right)
\left(\Omega_{1}^{2}(z\gg L_{abs})+\Omega_{2}^{2}(z\gg L_{abs})
+\Omega_{3}^{2}(z\gg L_{abs})\right)\right]^{-1/2}.
\label{eq:D-2}
\end{align}

\section{The multi-level system}

\begin{figure}
\includegraphics[width=0.5\columnwidth]{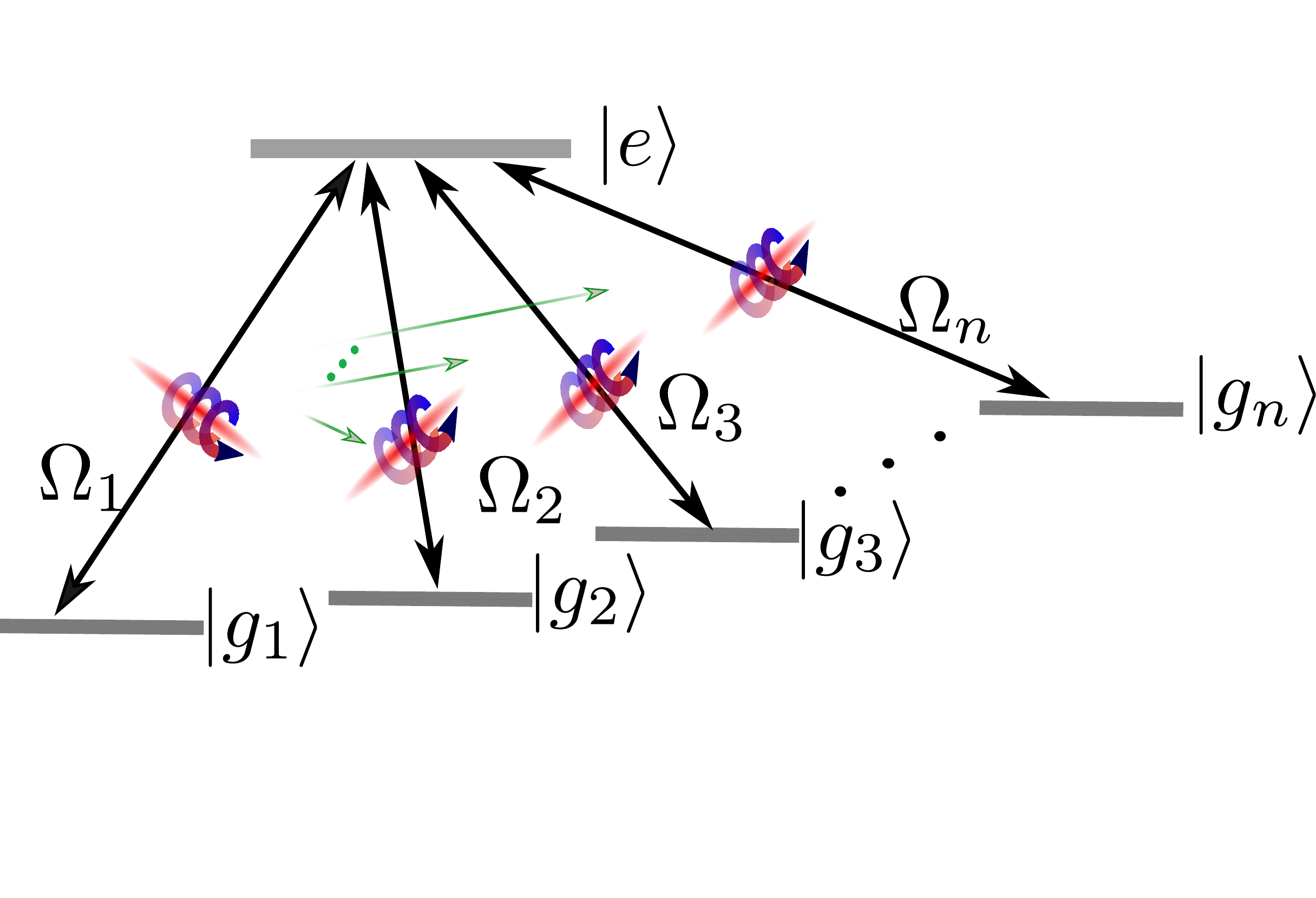}
\caption{Schematic diagram of the multi-level quantum system containing an
upper state $|e\rangle$ and lower levels $|g_{1}\rangle$, $|g_{2}\rangle$...
$|g_{n}\rangle$ interacting with Rabi-frequencies $\Omega_{1}$,
$\Omega_{2}$... $\Omega_{n}$.}
\label{fig:5}
\end{figure}

Let us now extend our model by considering the propagation of $n$-component
light pulses through an $(n+1)$-state atomic medium with $n$ lower
atomic states and one excited state shown in Fig.~\ref{fig:5}. Denoting
the excited state by $|e\rangle$, the lower levels by $|g_{1}\rangle$,
$|g_{2}\rangle$,..., $|g_{n}\rangle$ and the Rabi-frequency of laser
pulses by $\Omega_{m}$($m=1,2,...,n)$, the interaction Hamiltonian
for such a multi-level atom reads
\begin{equation}
H_{\mathrm{M}}=\sum_{m=1}^{n}\Omega_{m}|g_{m}\rangle\langle e|+\mathrm{H.c.}\,.\label{eq:33}
\end{equation}

The MBEs describing the dynamics of the system are given by
\begin{equation}
\dot{\rho}_{g_{m}e}=i(\delta_{m}+i\gamma_{eg_{m}})\rho_{g_{m}e}-i\Omega_{m}(\rho_{ee}-\rho_{g_{m}g_{m}})+i\sum_{j=1;j\neq m}^{n}\Omega_{j}\rho_{g_{m}g_{j}}\,.\label{eq:34}
\end{equation}
and
\begin{equation}
\frac{\partial\Omega_{m}}{\partial z}+c^{-1}\frac{\partial\Omega_{m}}{\partial t}=i\frac{\alpha_{m}\gamma_{eg_{m}}}{2L}\rho_{g_{m}e}\,(m=1,2,..n),\label{eq:35}
\end{equation}
with $m=1,2,...,n$. As before, the diffraction terms are ignored
in the Maxwell equations (\ref{eq:35}).

The atoms comprising the system are initially in the superposition
of $n$ ground states:
\begin{equation}
|\psi(0)\rangle=\sum_{m=1}^{n}c_{m}|g_{m}\rangle.\label{eq:atom-state3}
\end{equation}
Assuming the atom-light interaction to be sufficiently weak, $|\Omega_{j}|\ll\gamma_{eg_{j}}$,
one can approximate $\rho_{ee}\approx0$, $\rho_{g_{s}g_{s}}\approx|c_{s}|^{2}$,
$\rho_{g_{s}g_{t}}\approx c_{s}c_{t}^{*}$ to the first order in all
laser fields, giving
\begin{equation}
\rho_{g_{m}e}=-\frac{\sum_{j=1}^{n}c_{m}c_{j}^{*}\Omega_{j}}{\delta_{m}+i\gamma_{eg_{m}}},\,\,m=1,2,...,n\label{eq:36}
\end{equation}

If all the laser pulses except the first one are zero at the entrance
($\Omega_{1}(0)=\Omega$ while $\Omega_{2}(0)=0$, $\Omega_{3}(0)=0$,...,
$\Omega_{n}(0)=0$), the solutions to the MBEs (\ref{eq:34}) and
(\ref{eq:35}) are \cite{PaspalakisPhysRevA2002multi}
\begin{align}
\Omega_{1}(z)=&\frac{\Omega}{X_{n}}(\beta_{1}|c_{1}|^{2}e^{-iX_{n}z}+\sum_{N=2}^{n}\beta_{N}|c_{N}|^{2}),\label{eq:37-1}\\
\Omega_{N}(z)=&\frac{\Omega}{X_{n}}c_{1}^{*}c_{N}\beta_{N}(e^{-iX_{n}z}-1),\,\,(N=2,..,n),\label{eq:38-1}
\end{align}
with
\begin{equation}
X_{n}=\sum_{m=1}^{n}\beta_{m}|c_{m}|^{2}.\label{eq:xn}
\end{equation}
From Eqs.~(\ref{eq:15}) and (\ref{eq:16}) it follows then that
if the first light pulse photons carry an OAM of $\hbar l$ along
the propagation direction, $n-1$ optical vortices are generated in
the medium with the same vorticity as the first laser beam $\Omega_{1}$.
In the vicinity of the vortex core the generated vortex beams look
like a LG beam with their intensity vanishing at the core $r\rightarrow0$. 

Calling on Eqs.~(\ref{eq:15}) and (\ref{eq:16}), Eqs.~(\ref{eq:37-1})
and (\ref{eq:38-1}) provide the following solutions for the distances
exceeding the absorption length $z\gg L_{abs}$
\begin{align}
\Omega_{1}(z\gg L_{abs})=&\varepsilon(\frac{r}{w})^{|l|}e^{-r^{2}/w^{2}}(1-|c_{1}|^{2})e^{il\phi},\label{eq:39}\\
\Omega_{N}(z\gg L_{abs})=&-\varepsilon(\frac{r}{w})^{|l|}e^{-r^{2}/w^{2}}c_{1}^{*}c_{N}e^{il\phi},\label{eq:40}
\end{align}
where we have assumed that $\alpha_{1}=\alpha_{2}=...=\alpha_{n}=\alpha$,
$\gamma_{eg_{1}}=\gamma_{eg_{2}}=...=\gamma_{eg_{n}}=\gamma$, and
$\delta_{1}=\delta_{2}=...=\delta_{n}=0$. Equations (\ref{eq:39})-(\ref{eq:40})
demonstrate the lossless propagation of the $n$-component optical
vortices because for $z\gg L_{abs}$ the multi-level model goes to
a linear superposition of $n-1$ dark states. 

\section{Composite vortices}

Let us next consider a situation where the $\Lambda$ scheme is initially
prepared in a superposition state given by Eqs.~(\ref{eq:atom-state1}),
but both fields $\Omega_{1}$ and $\Omega_{2}$ are incident on the
medium. With the initial conditions where both incident fields are
the vortex beams $\Omega_{1}(0)=\Omega_{10}=\varepsilon_{1}(\frac{r}{w})^{|l_{1}|}e^{-r^{2}/w^{2}}e^{il_{1}\phi}$
and $\Omega_{2}(0)=\Omega_{20}=\varepsilon_{2}(\frac{r}{w})^{|l_{2}|}e^{-r^{2}/w^{2}}e^{il_{2}\phi}$,
the solutions to the Eqs.~(\ref{eq:8}) and (\ref{eq:9}) take the
form
\begin{equation}
\Omega_{1}(z)=\frac{1}{X_{2}}\left[\varepsilon_{1}(\frac{r}{w})^{|l_{1}|}e^{-r^{2}/w^{2}}\left(\beta_{1}|c_{1}|^{2}e^{-iX_{2}z}+\beta_{2}|c_{2}|^{2}\right)e^{il_{1}\phi}+\varepsilon_{2}(\frac{r}{w})^{|l_{2}|}e^{-r^{2}/w^{2}}c_{1}c_{2}^{*}\beta_{1}\left(e^{-iX_{2}z}-1\right)e^{il_{2}\phi}\right],\label{eq:41-1}
\end{equation}

\begin{equation}
\Omega_{2}(z)=\frac{1}{X_{2}}\left[\varepsilon_{1}(\frac{r}{w})^{|l_{1}|}e^{-r^{2}/w^{2}}c_{1}^{*}c_{2}\beta_{2}\left(e^{-iX_{2}z}-1\right)e^{il_{1}\phi}+\varepsilon_{2}(\frac{r}{w})^{|l_{2}|}e^{-r^{2}/w^{2}}\left(|c_{2}|^{2}\beta_{2}e^{-iX_{2}z}+|c_{1}|^{2}\beta_{1}\right)e^{il_{2}\phi}\right].\label{eq:42-1}
\end{equation}
In this way by applying two incident vortex beams $\Omega_{1}(0)$
and $\Omega_{2}(0)$ one produces two new beams $\Omega_{1}(z)$ and
$\Omega_{2}(z)$ which may contain different vortices depending on
the relative amplitude and phase of the incident beams. 

\begin{figure}
\includegraphics[width=0.3\columnwidth]{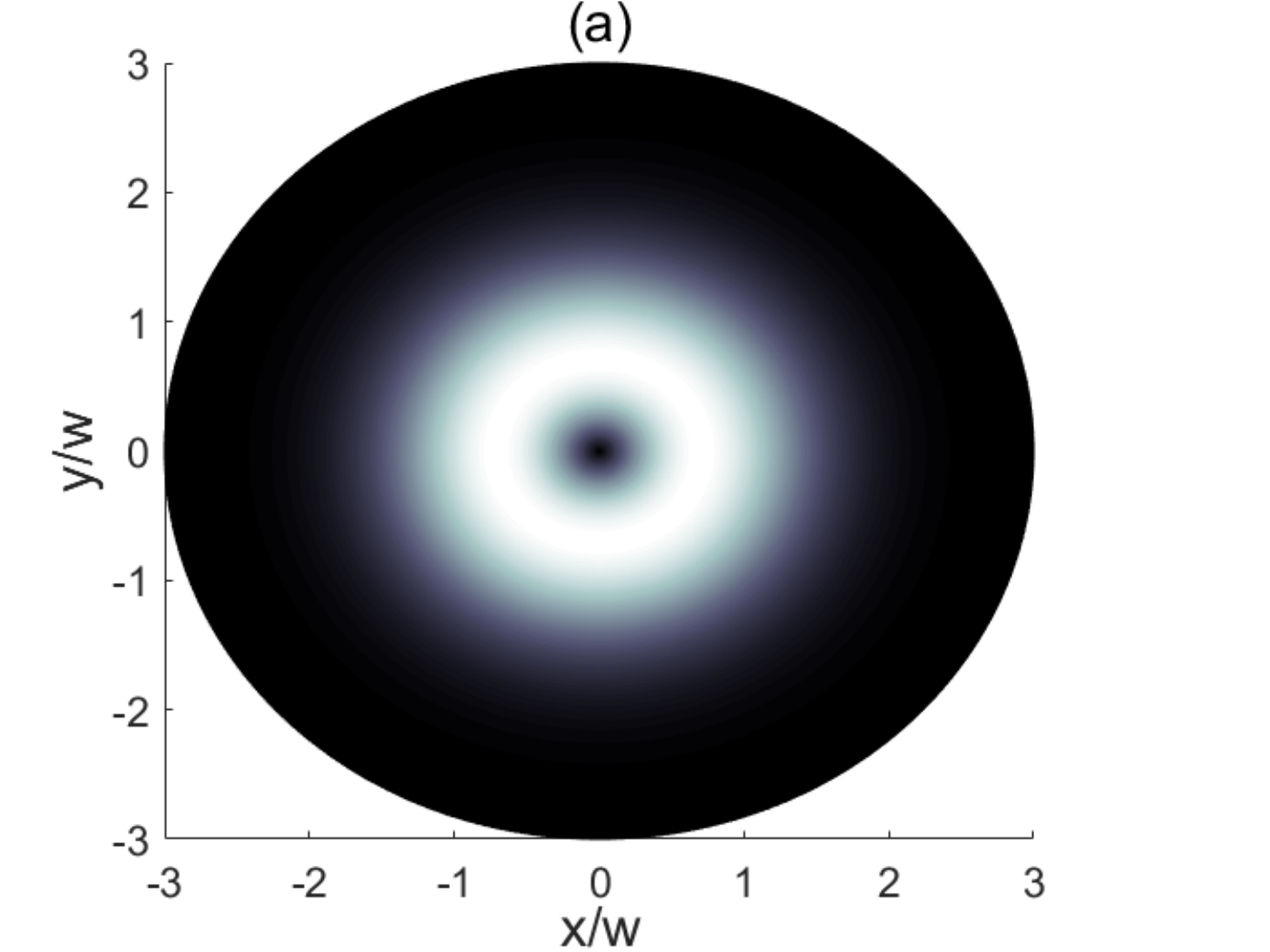}\includegraphics[width=0.3\columnwidth]{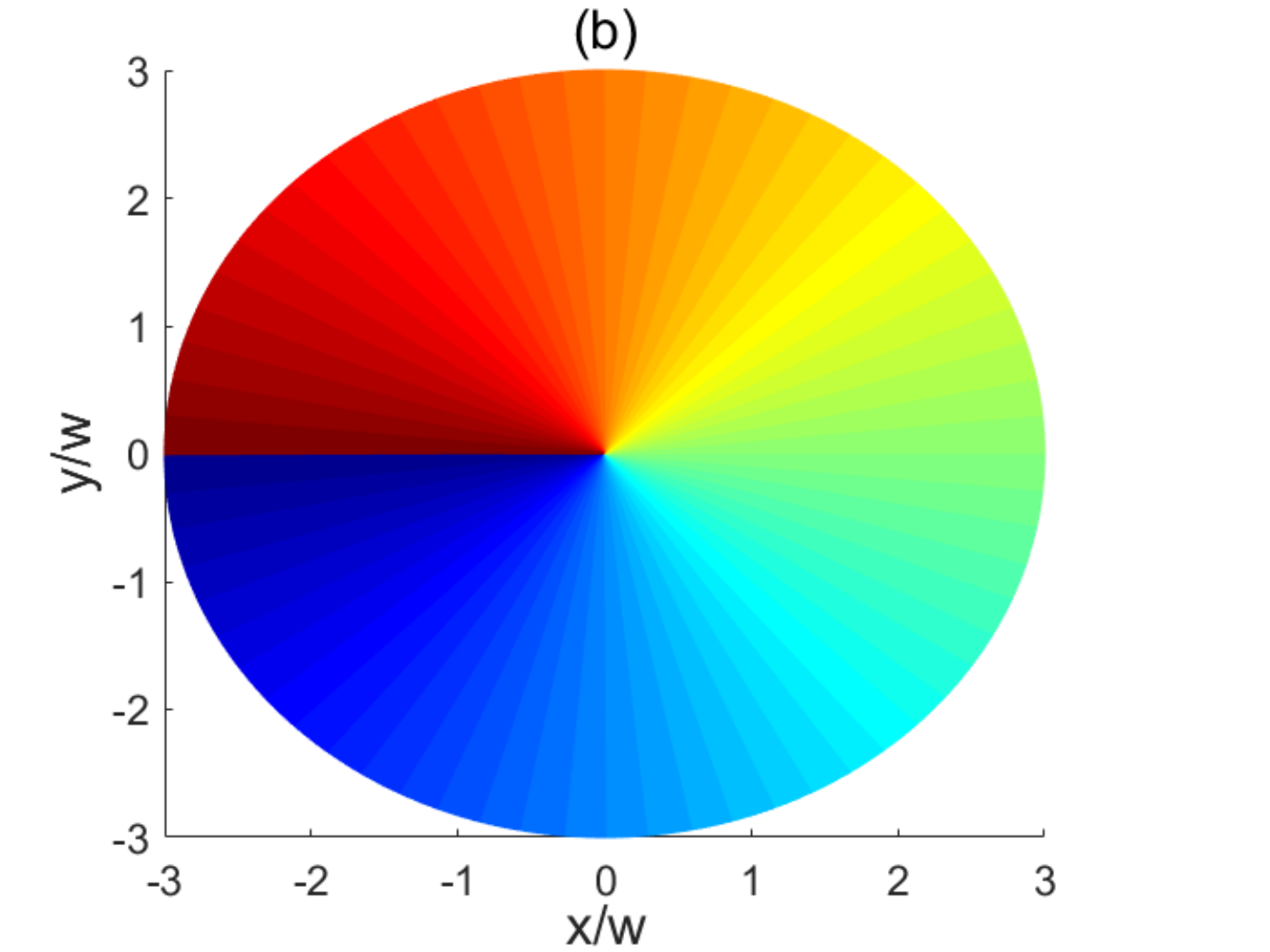}

\includegraphics[width=0.3\columnwidth]{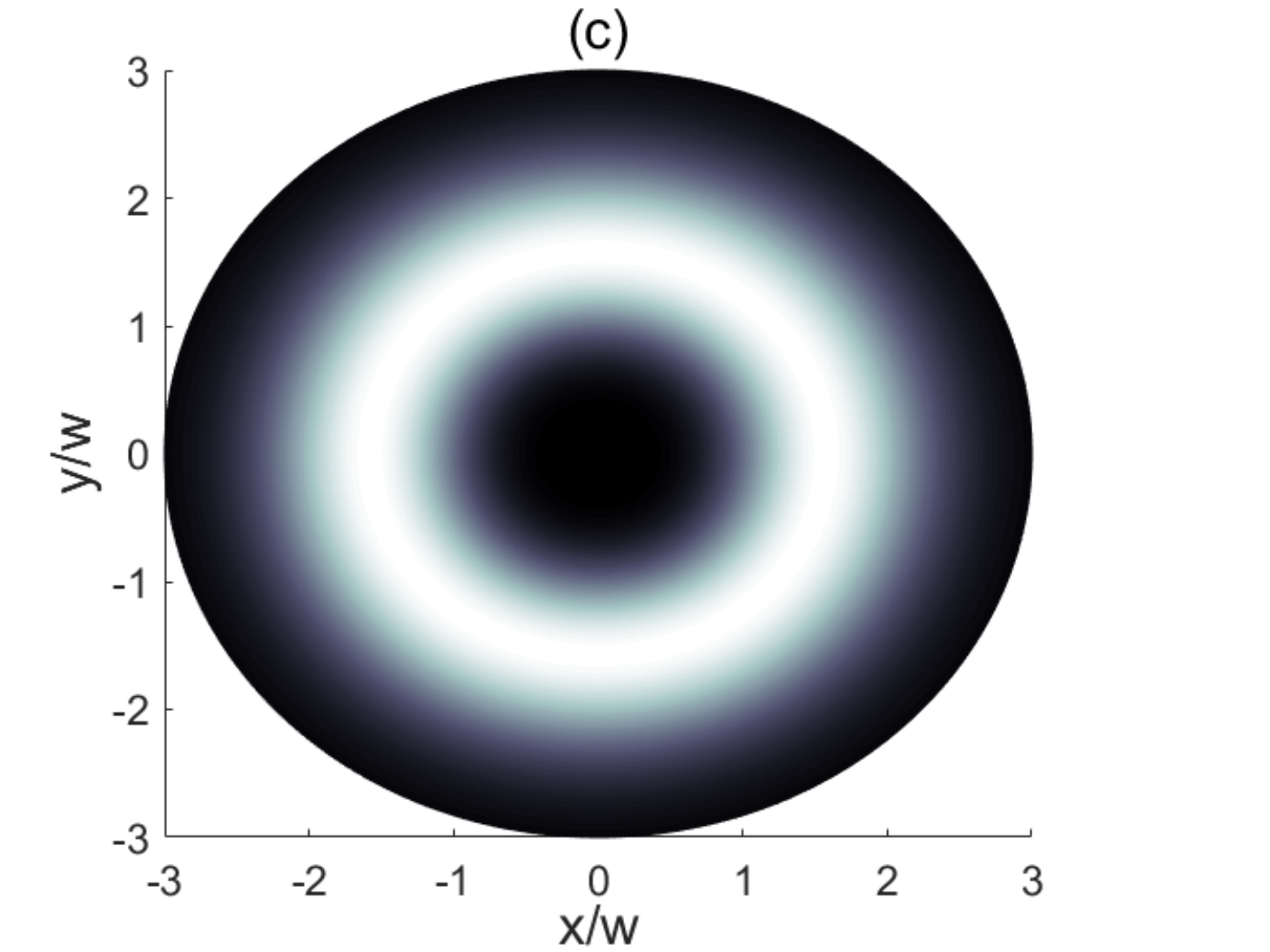}\includegraphics[width=0.3\columnwidth]{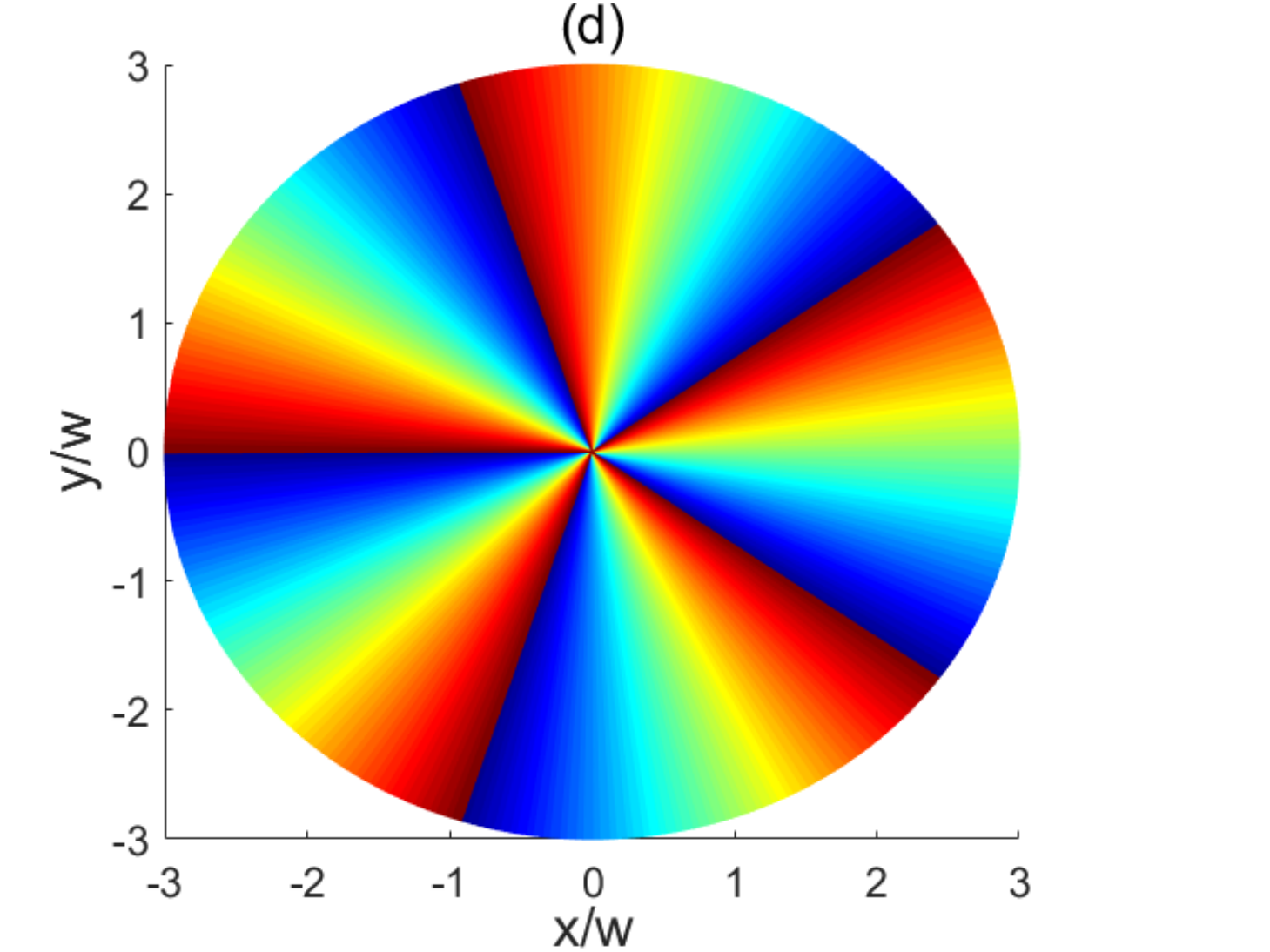} 

\includegraphics[width=0.3\columnwidth]{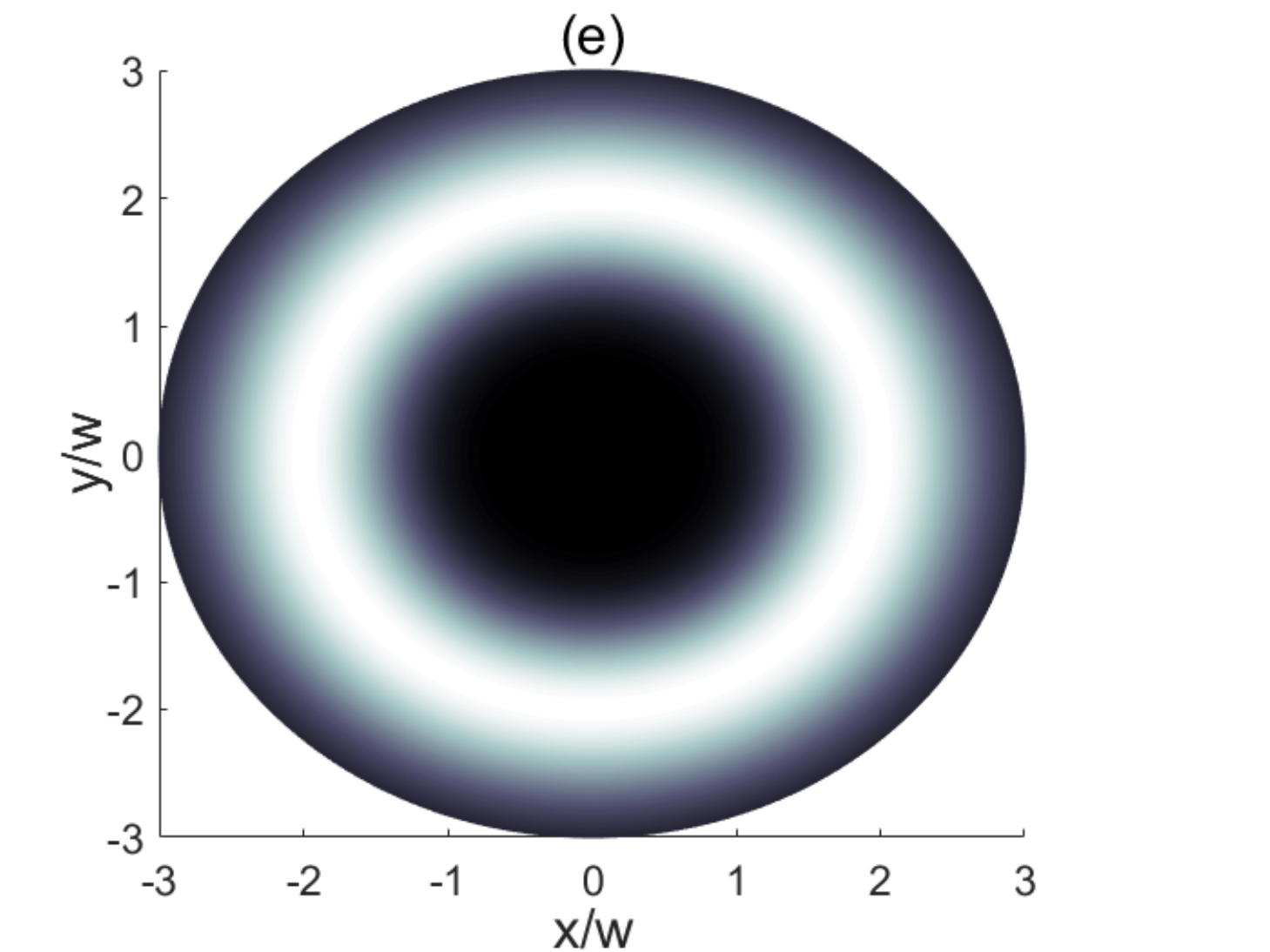}\includegraphics[width=0.3\columnwidth]{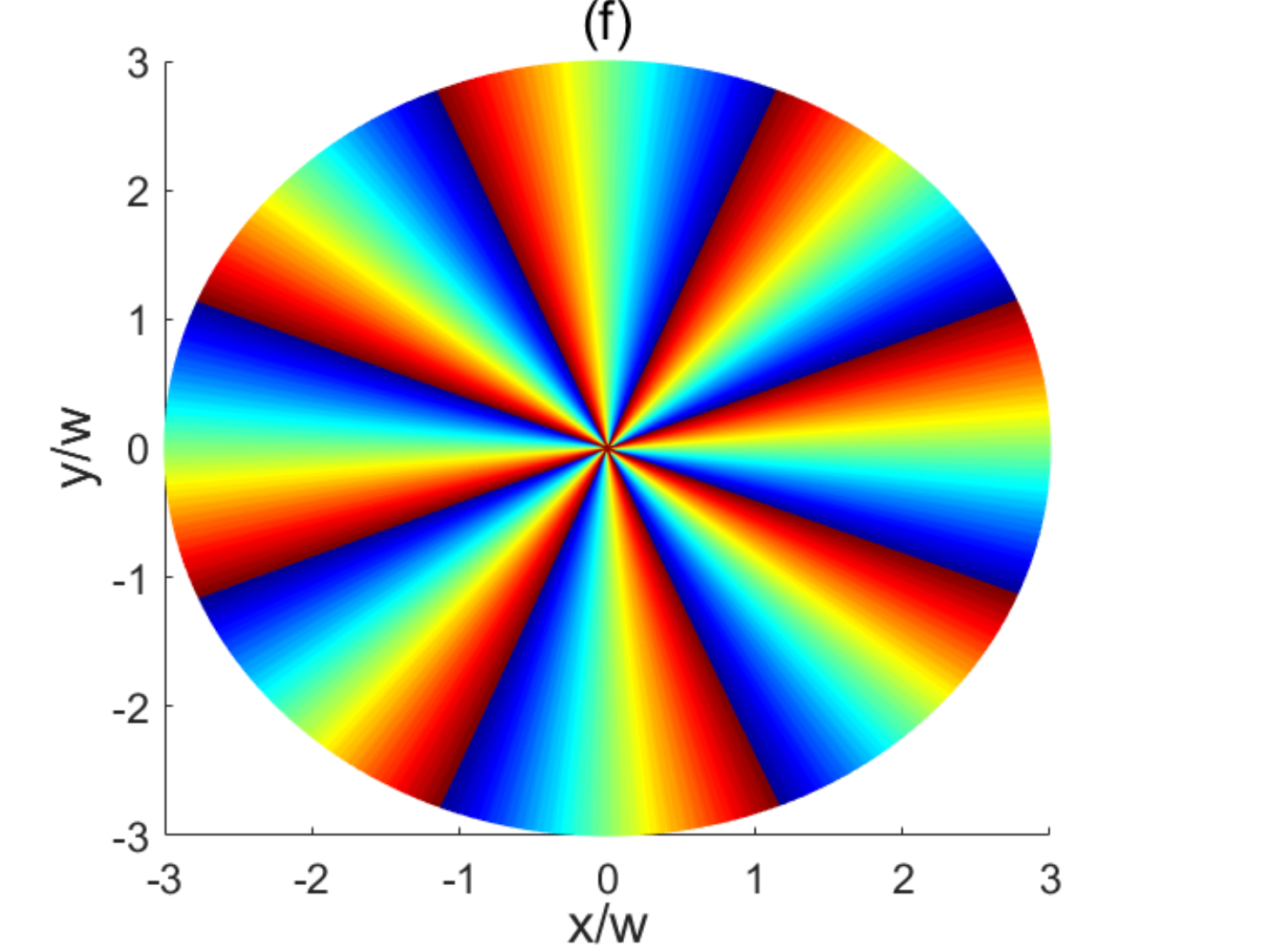}

\caption{(a), (c), (e) Intensity distributions in arbitrary units as well as
(b), (d), (f) the corresponding helical phase patterns of the beam
$\Omega_{1}(z)$ defined by Eq.~(\ref{eq:41-1}) generated by combining
two vortex beams with vorticities  (a), (b) $l_{1}=l_{2}=1$, (c),
(d) $l_{1}=l_{2}=5$ and (e), (f) $l_{1}=l_{2}=8$. Here, the parameters
are $z=L/2$, $c_{1}=c_{2}=\frac{1}{\sqrt{2}}$, $\delta_{1}=\delta_{2}=0$,
$\varepsilon_{1}=\varepsilon_{2}$ and $\alpha=20$. The intensity
distribution and phase pattern of the field $\Omega_{2}(z)$ are identical
to the intensity distribution and phase pattern of the field $\Omega_{1}(z)$
shown in this figure.}
\label{fig:6}
\end{figure}

\begin{figure}
\includegraphics[width=0.3\columnwidth]{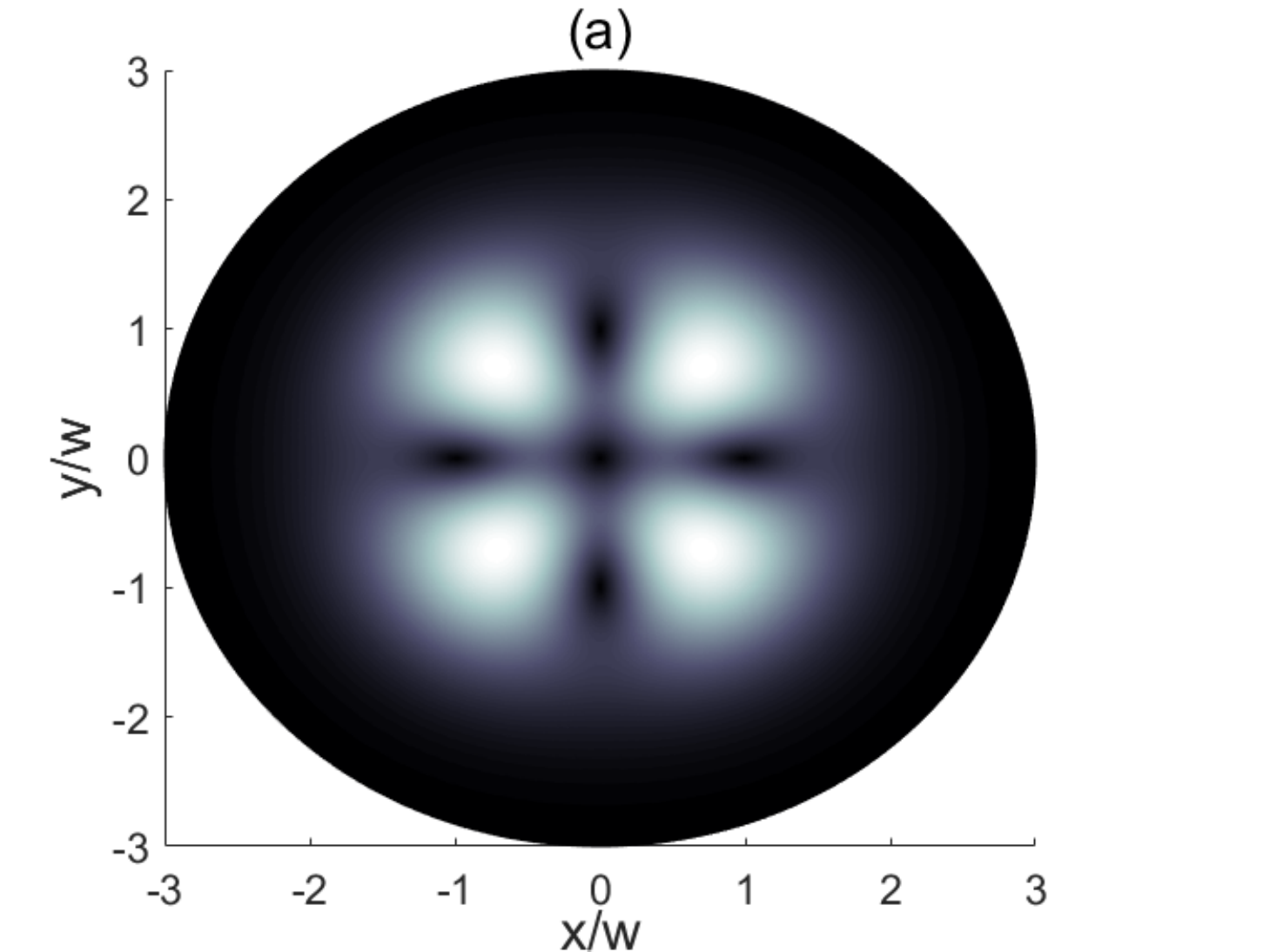}\includegraphics[width=0.3\columnwidth]{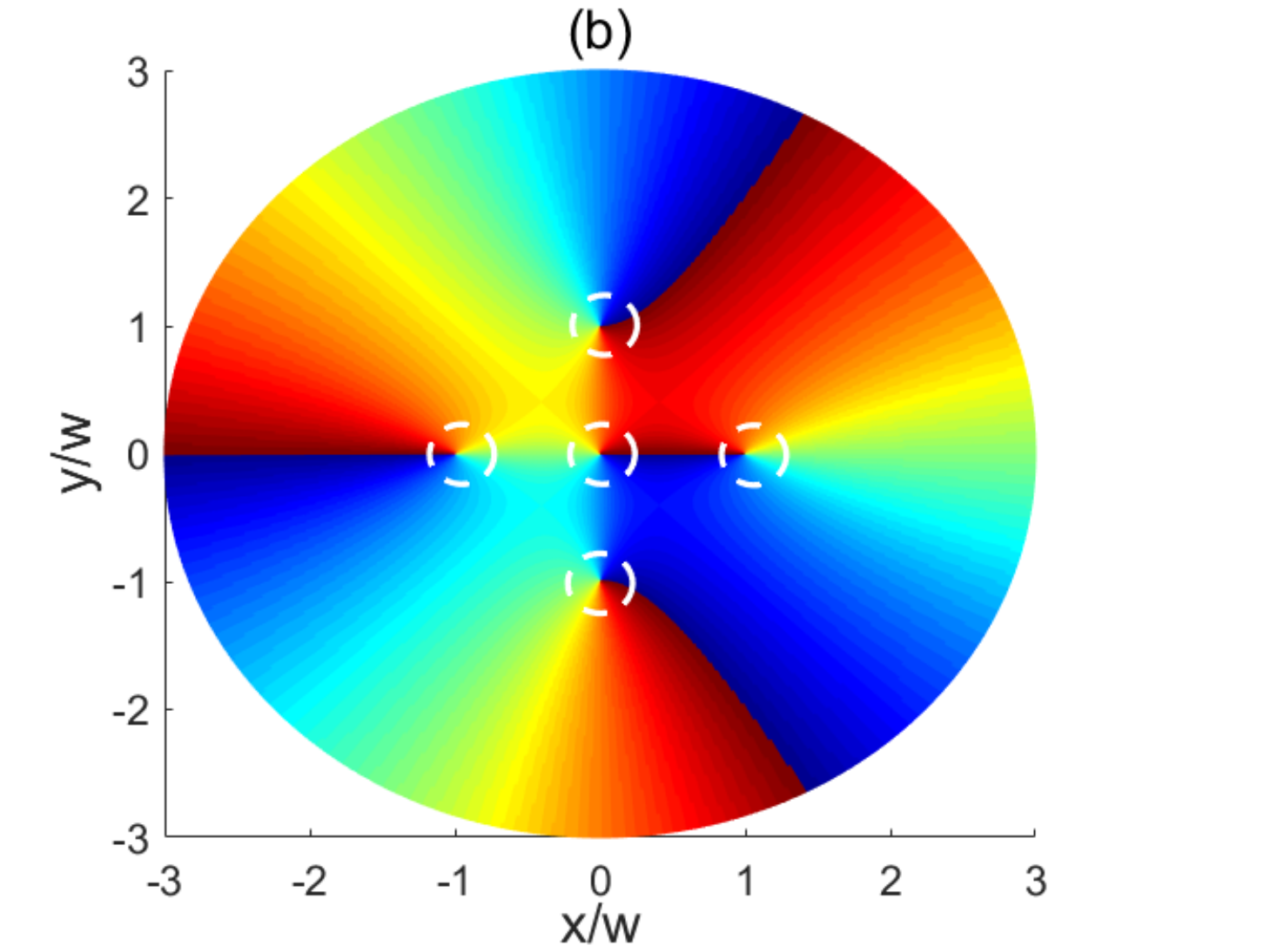}

\includegraphics[width=0.3\columnwidth]{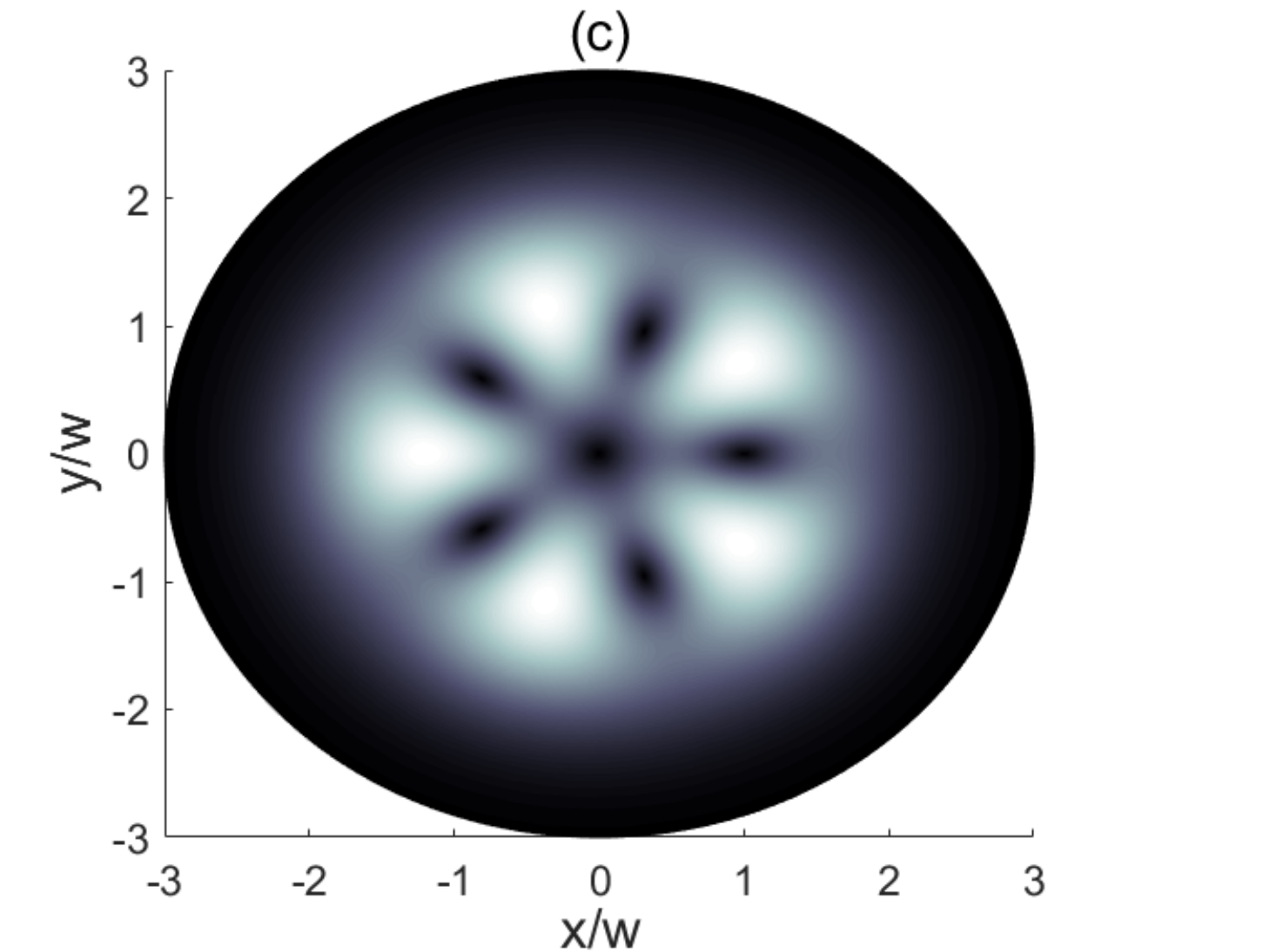}\includegraphics[width=0.3\columnwidth]{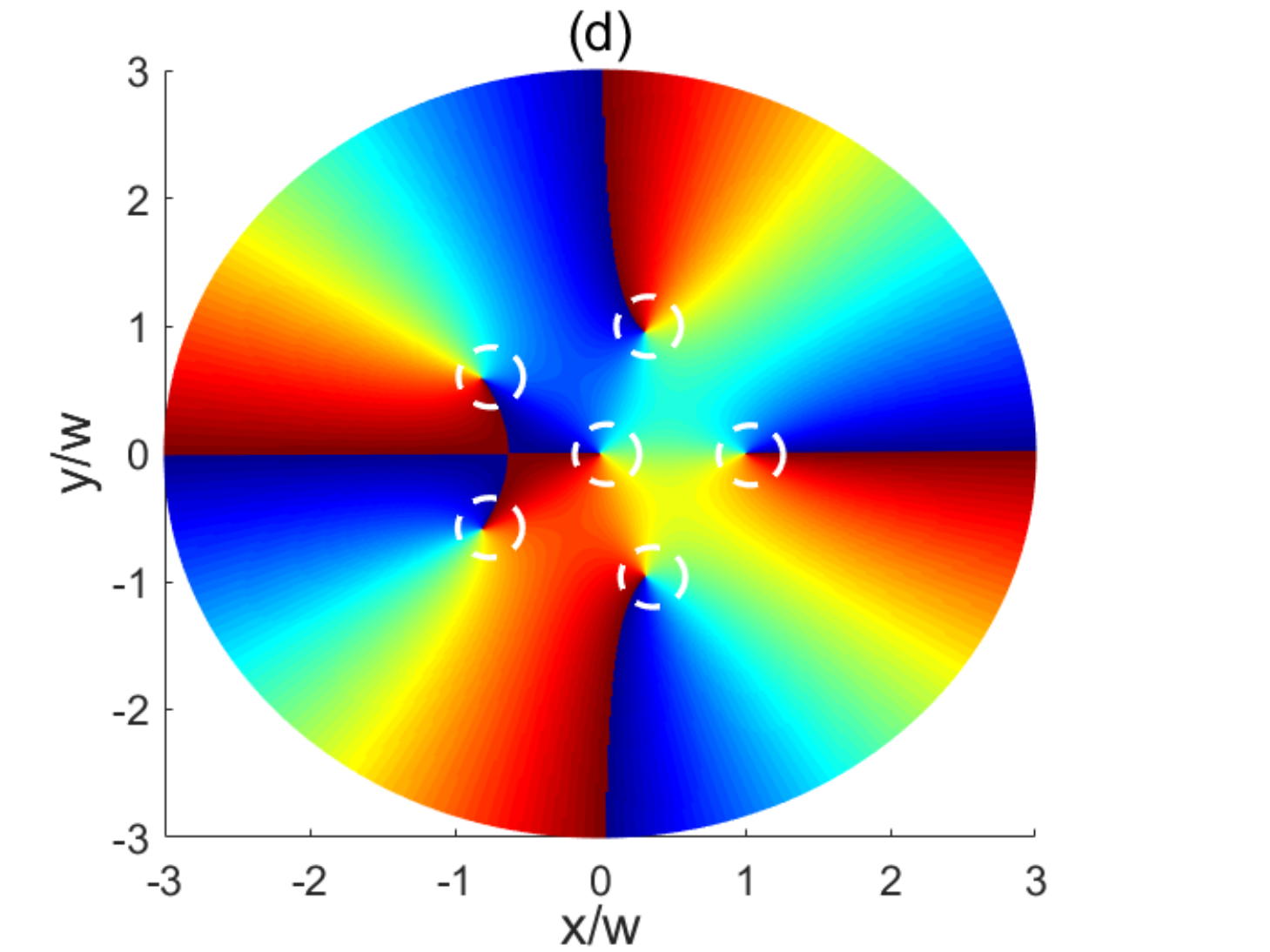}

\includegraphics[width=0.3\columnwidth]{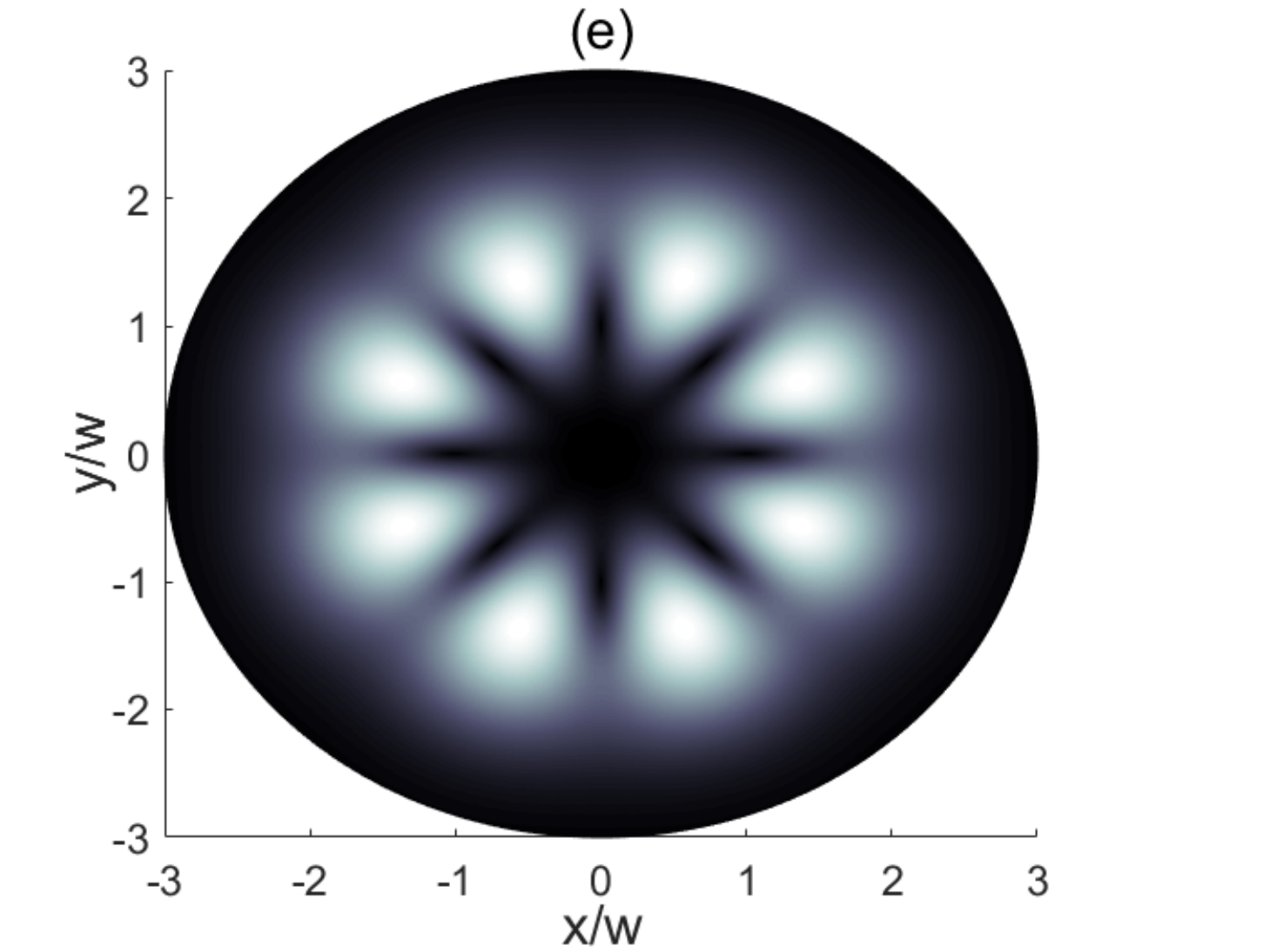}\includegraphics[width=0.3\columnwidth]{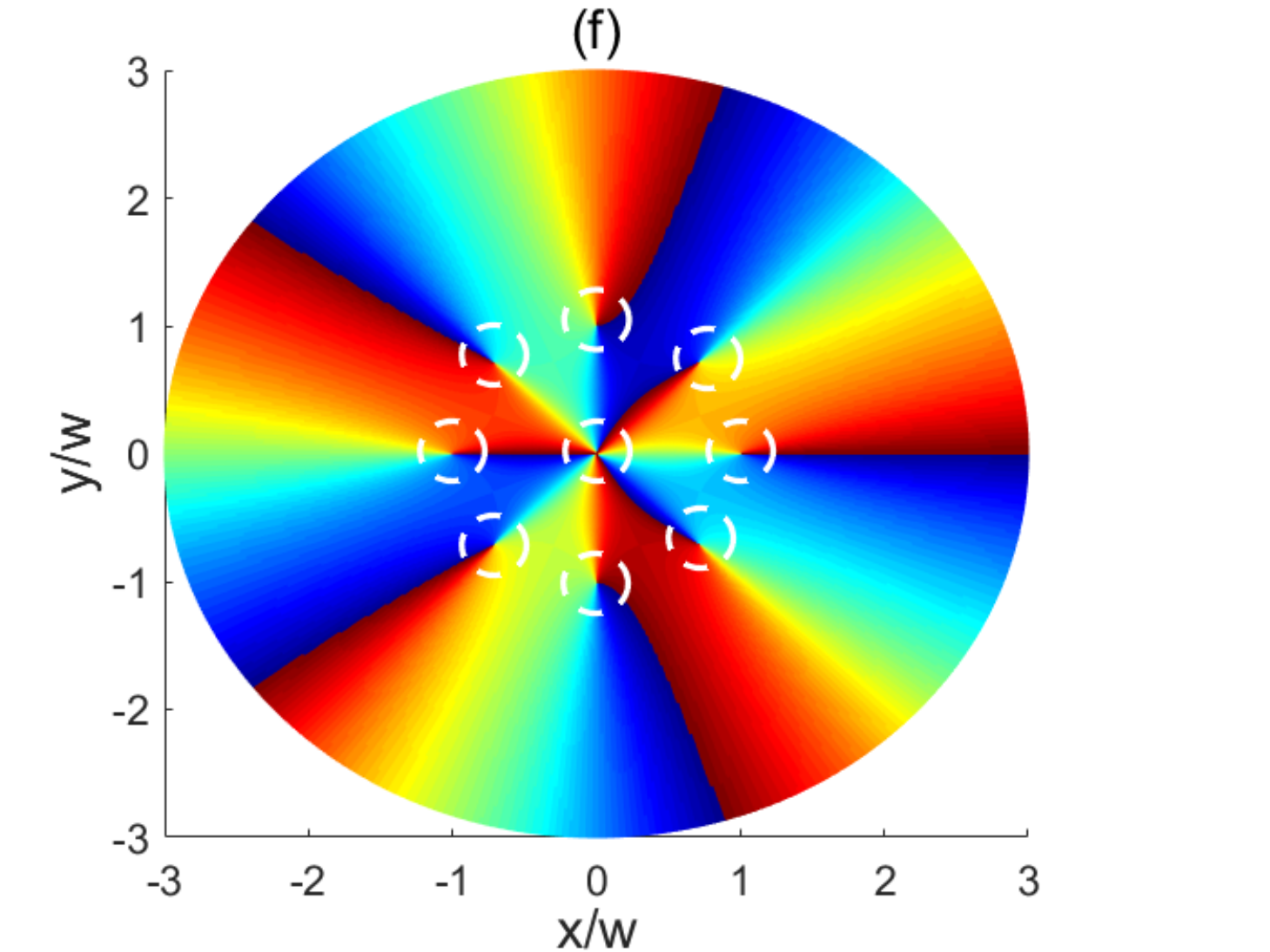} 

\caption{(a), (c), (e) Intensity distributions in arbitrary units as well as
the (b), (d), (f) corresponding helical phase patterns of the beam
$\Omega_{1}(z)$ defined by Eq.~(\ref{eq:41-1}) generated by combining
two vortex beams with vorticities (a), (b) $l_{1}=1$, $l_{2}=-3$,
(c), (d) $l_{1}=-1$, $l_{2}=4$ and (e), (f) $l_{1}=3$, $l_{2}=-5$
. The selected parameters are the same as in Fig.~\ref{fig:6}. The
white dash lines in the phase patterns show the position of vortices.
The intensity distribution and phase pattern of the field $\Omega_{2}(z)$
are identical to the intensity distribution and phase pattern of the
field $\Omega_{1}(z)$ shown in this figure.}
\label{fig:7}
\end{figure}

\begin{figure}
\includegraphics[width=0.3\columnwidth]{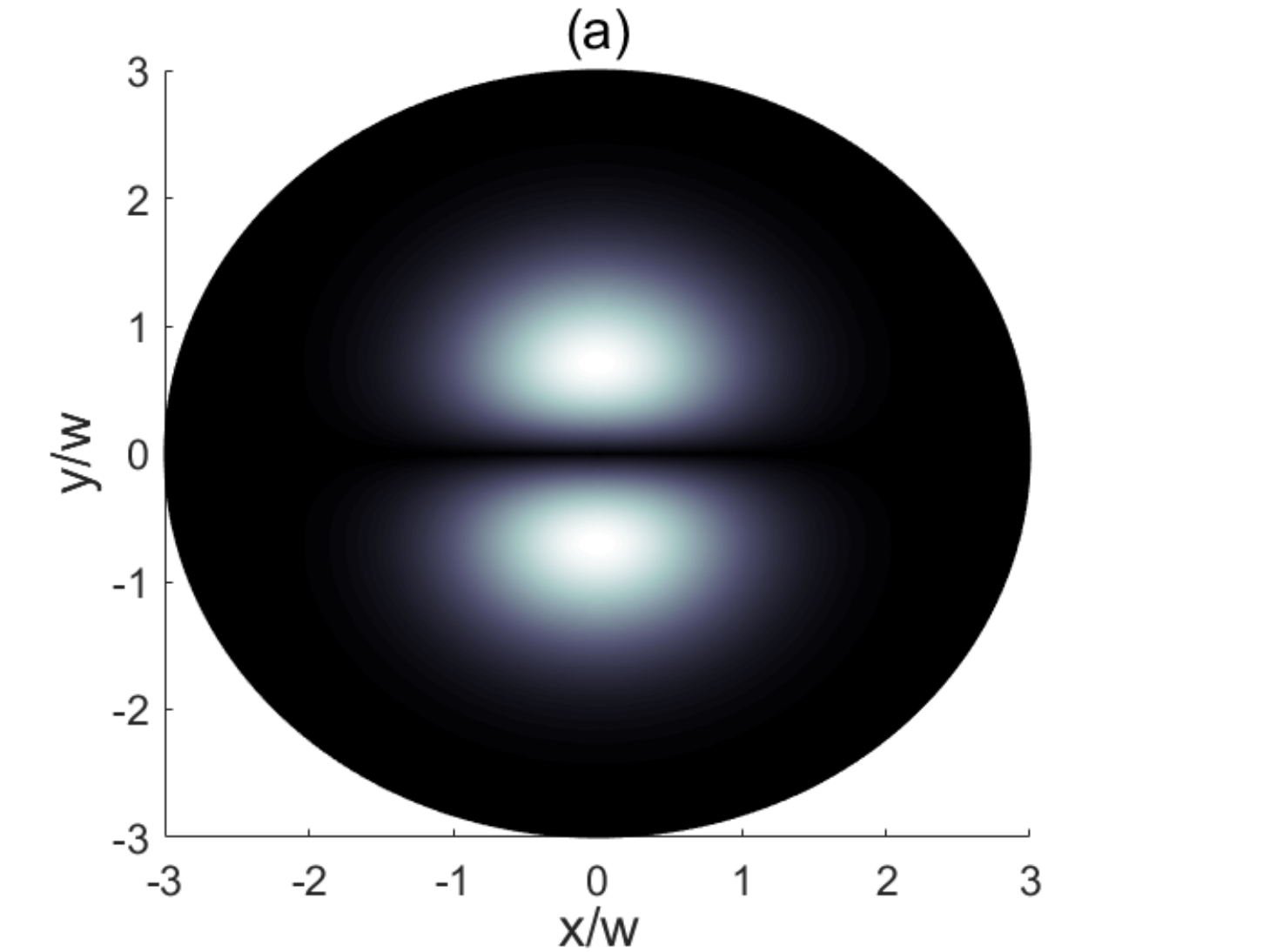}\includegraphics[width=0.3\columnwidth]{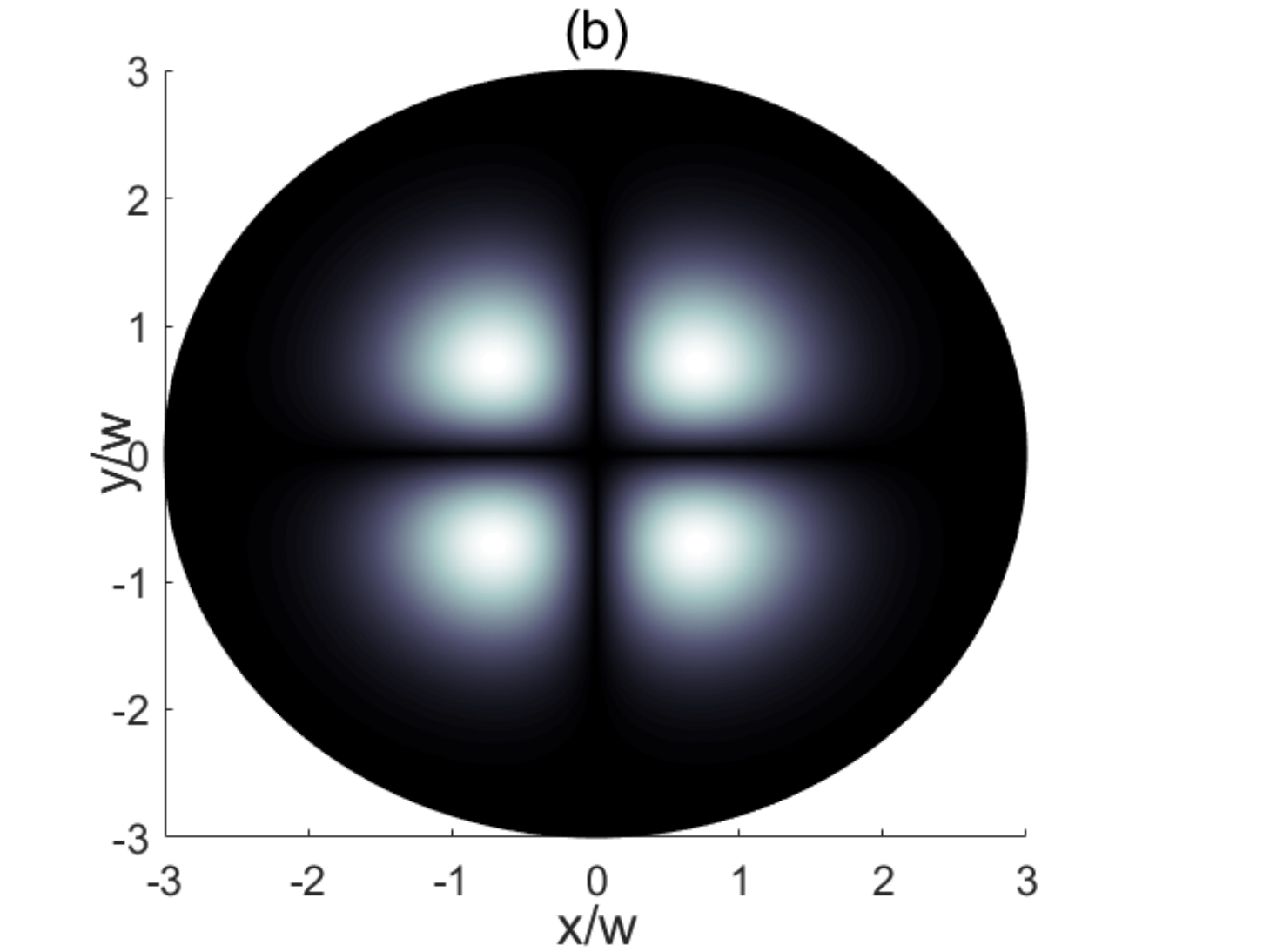}

\includegraphics[width=0.3\columnwidth]{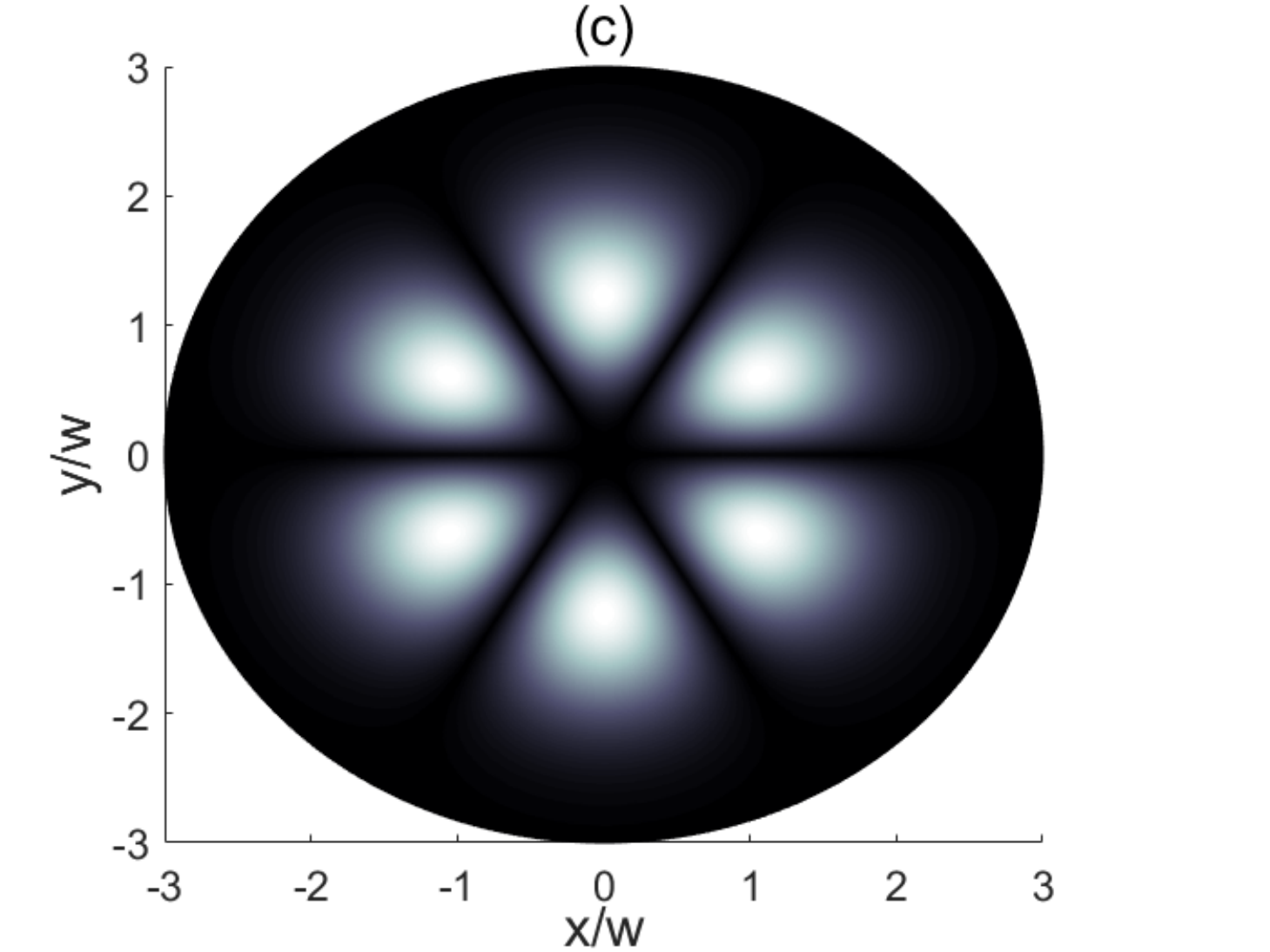}\includegraphics[width=0.3\columnwidth]{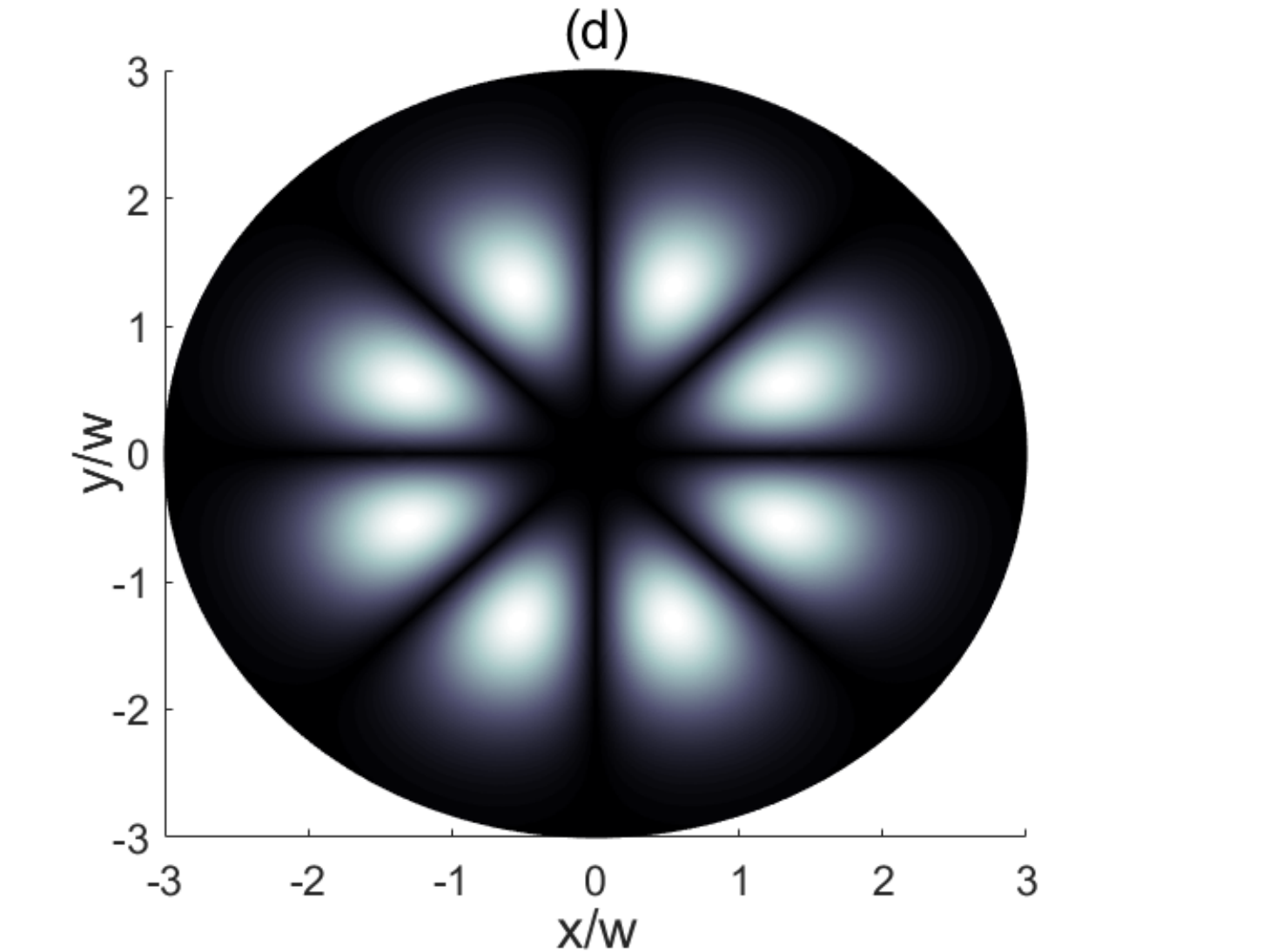}

\caption{Intensity distributions in arbitrary units for the beam $\Omega_{1}(z)$
defined by Eq.~(\ref{eq:41-1}) generated by combining two vortex
beams with vorticities (a) $l_{1}=1$, $l_{2}=-1$, (b) $l_{1}=2$,
$l_{2}=-2$, (c) $l_{1}=3$, $l_{2}=-3$, and (c) $l_{1}=4$, $l_{2}=-4$.
The selected parameters are the same as Fig.~\ref{fig:6}. The intensity
distribution of the field $\Omega_{2}(z)$ is identical to the intensity
distribution of the field $\Omega_{1}(z)$ shown in this figure.}
\label{fig:8}
\end{figure}

Various situations can appear for the beams created in this way. If
the winding numbers of the incident pulses are the same, $l_{1}=l_{2}=l$,
the resulting beams $\Omega_{1}(z)$ and $\Omega_{2}(z)$ have the
same vorticity $l$, and the vortex width increases with increasing
the winding number $l$, as illustrated in Figs.~\ref{fig:6}(a),\ref{fig:6}(c),\ref{fig:6}(e).
Figures~\ref{fig:6}(b),\ref{fig:6}(d),\ref{fig:6}(f)) show the
corresponding phase profile of the beams. If $|l_{1}|<|l_{2}|$, the
resulting composite twisted beam contains a vortex of charge $l_{1}$
located at the beam center which is surrounded by $|l_{1}-l_{2}|$
peripheral vortices (Fig.~\ref{fig:7}). In this case, two light
vortices with different winding numbers $l_{1}$ and $l_{2}$ around
same axis result in formation of vortices with shifted axes. In particular,
for $l_{1}=-l_{2}=l$ we superimpose two optical vortices with opposite
topological charges and equal intensity, and the azimuthal dependence
is given by $e^{il\phi}+e^{-il\phi}=2\cos(l\phi)$. This corresponds
to the flower-like \textquotedblleft petals\textquotedblright{} intensity
structures demonstrated in Fig.~\ref{fig:8}. Note that such a flower-like
structure is not called as vortex, although it has a zero intensity
at the center \cite{Franke-Arnold2007OE,BaumannOE2009}.


Let us know assume $\alpha_{1}=\alpha_{2}=\alpha$, $\gamma_{eg_{1}}=\gamma_{eg_{2}}=\gamma$,
and $\delta_{1}=\delta_{2}=0$. At propagation distances exceeding
the absorption length $z\gg L_{abs}$ all exponential terms vanish
in Eqs.~(\ref{eq:41-1}) and (\ref{eq:42-1}) and we get
\begin{equation}
\Omega_{1}(z\gg L_{abs})=e^{-r^{2}/w^{2}}\left[|c_{2}|^{2}\varepsilon_{1}(\frac{r}{w})^{|l_{1}|}e^{il_{1}\phi}-c_{1}c_{2}^{*}\varepsilon_{2}(\frac{r}{w})^{|l_{2}|}e^{il_{2}\phi}\right],\label{eq:16-1-1}
\end{equation}

\begin{equation}
\Omega_{2}(z\gg L_{abs})=e^{-r^{2}/w^{2}}\left[-c_{1}^{*}c_{2}\varepsilon_{1}(\frac{r}{w})^{|l_{1}|}e^{il_{1}\phi}+|c_{1}|^{2}\varepsilon_{2}(\frac{r}{w})^{|l_{2}|}e^{il_{2}\phi}\right].\label{eq:16-2-1}
\end{equation}
Therefore if the two incident vortex fields are nonzero, for $z\gg L_{abs}$
both vortex beams experience no absorption as the atoms comprising
the medium are converted to their dark states defined by Eq.~(\ref{eq:D}).
It is noteworthy that there is an even more favorable scenario for
the lossless propagation of both vortex beams. Assuming that $\Omega_{1}(0)=\Omega_{2}(0)=\Omega=\varepsilon(\frac{r}{w})^{|l|}e^{-r^{2}/w^{2}}e^{il\phi}$
and choosing the values of $c_{1}$ and $c_{2}$ such that $c_{1}=-c_{2}=\frac{1}{\sqrt{2}}$,
one arrives at $\Omega_{1}(z)=\Omega_{2}(z)=\Omega(z)$. Under this
condition the atoms are in their dark states from the very beginning,
the medium becomes completely transparent to both vortex beams, and
the fields propagate without losses as in free space. Such an analysis
for generation and propagation of composite optical vortices can be
extended to the $n+1$-level schemes when all $n$ laser fields are
present at the entrance to the atomic medium. 

\section{Concluding Remarks\label{sec:concl}}

We have analyzed the propagation dynamics of two (three) component
laser pulses with OAM interacting with atoms in the $\Lambda$ (tripod)
atom-light coupling schemes. The quantum system is initially prepared
in a coherent superposition of two (three) lower levels. If a vortex
beam acts on one transition of the $\Lambda$ (tripod) system, an
extra light beams can be nonlinearly generated with the same OAM number
as the initially injected structured light. We have also extended
the analysis to a $n+1$-level phaseonium medium for the $n$-component
generation of the twisted light beams. The lossless propagation of
generated vortex light beams has been also considered. It has been
shown that at the propagation distances exceeding the absorption length
the system goes to a linear superposition of $n-1$ dark states leading
to the transparency of the medium to the $n$-component optical vortices.

It has been recently shown that a double-$\Lambda$ scheme can be
employed for the exchange of optical vortices based on EIT \cite{Hamedi-PhysRevA-2018}.
In the double-$\Lambda$ scheme there should be two additional control
lasers of larger intensity to assure the exchange of optical vortices.
On the other hand, in the current proposal one does not need the strong
atom-light interaction as we are dealing with small intensities ($|\Omega_{i}|\ll\gamma_{eg_{i}}$).
It is only needed that a medium is initially coherently prepared in
a superposition of atomic lower levels. The losses in both schemes
are similar and take place mostly at beginning of the medium within
the absorption length. The losses disappear when the light pulses
propagate deeper through the medium. 

We have also considered a situation where both vortex beams $\Omega_{1}$
and $\Omega_{2}$ are present at beginning of the medium of the $\Lambda$-type
atoms. When the two vortex beams are incident on the medium, they
can create two composite beams with new vortices. Different cases
for the appearance of composite vortices have been explored, and the
situations for absorptionless propagation of composite vortices are
discussed. We have also extended the model for generation of composite
optical vortices to the $n+1$-level structures. 

The coherent superposition of the ground states employed in this paper
can be realized experimentally using the fractional or partial STIRAP
in which only a controlled fraction of the population is transferred
to the target state \cite{Vitanov-RevModPhys-2017}. The creation
of a quantum superposition of metastable states out of a single initial
state in a robust and controlled way has been shown to be possible
in a four-state system by using a sequence of three pulses \cite{Unanyan98OC,Unanyan99PRA}.
Such a technique is based on the existence of two degenerate dark
states and their interaction. The mixing of the dark states can be
controlled by changing the relative delay of the pulses, and thus
an arbitrary superposition state can be generated. Such a method for
creation of coherent superpositions can be generalized to $N$ level
schemes. 

The $\Lambda$ (tripod) level scheme containing two (three) ground
states and one excited state may implemented experimentally for example
using the $^{87}Rb$ atoms. The excited level $|e\rangle$ can then
correspond to the $|5P_{1/2},F=1,m_{F}=0\rangle$ state. The lower
states $|g_{1}\rangle$ and $|g_{2}\rangle$ (and $|g_{3}\rangle$)
can be attributed to the $|5S_{1/2},F=1,m_{F}=1\rangle$ and $|5S_{1/2},F=1,m_{F}=-1\rangle$
(and $|5S_{1/2},F=2,m_{F}=0\rangle$) \cite{Si2009}. 
\begin{acknowledgments}
This research was funded by the European Social Fund under grant No.\ 09.3.3-LMT-K-712-01-0051.
H.R.H gratefully acknowledges professor Lorenzo Marrucci for useful
discussions on composite optical vortices. 
\end{acknowledgments}

\end{document}